\documentclass{aa}  

\usepackage{graphicx}
\usepackage{txfonts}
\usepackage{lipsum}
\usepackage{subcaption}         
\usepackage{lscape}             
\usepackage{placeins}           
\usepackage{xcolor} 
\usepackage{booktabs}
\usepackage{multirow}

\AtBeginDocument{%
  \nolinenumbers
  \renewcommand{\linenumbers}[1][]{}
}

\newcommand{\hi}{H{\small I}}
\newcommand{\oii}{[O{\small II}]}

\newcommand{\mgii}{Mg\,\textsc{ii}}

\newcommand{\mgiiaw}{Mg\,\textsc{ii}~$\lambda$2796}

\graphicspath{{./}{figures/}}

\begin{document}

\title{MusE GAs FLOw and Wind (MEGAFLOW) XIV:}
\subtitle{Background-Galaxy Absorption Reveals Kiloparsec-Scale Structure in the Cool Circumgalactic Medium}

%
%
%

   \author{Yucheng Guo\inst{1}\corrauth{yuchengg@asu.edu}        
        \and Nicolas F. Bouché\inst{2}
        \and Martin Wendt\inst{3}
        \and Timothy Heckman\inst{1,4}
        \and Joop Schaye\inst{5}
        \and Sanchayeeta Borthakur\inst{1}
        \and Johannes Zabl\inst{2}
        \and Maxime Cherrey\inst{2}
        \and Sowgat Muzahid \inst{6}
        \and Ismael Pessa\inst{7}
        \and Ramona Augustin\inst{7}
        \and Daria Kozlova\inst{7}
        }

   \institute{School of Earth \& Space Exploration, Arizona State University, 781 Terrace Mall, Tempe, AZ 85287, USA
   \and Univ Lyon, Univ Lyon1, Ens de Lyon, CNRS, Centre de Recherche Astrophysique de Lyon UMR5574, F-69230, Saint-Genis-Laval, France
   \and Institut f{\"u}r Physik und Astronomie, Universit{\"a}t Potsdam,Karl-Liebknecht-Str. 24/25, D-14476 Golm, Germany
   \and Department of Physics \& Astronomy, Johns Hopkins University, Bloomberg Centre, 3400 N. Charles Street, Baltimore, MD 21218, USA
   \and Leiden Observatory, Leiden University, PO Box 9513, NL-2300 RA Leiden, the Netherlands
   \and Inter-University Centre for Astronomy \& Astrophysics, Post Bag 04, Ganeshkhind, Pune 411007, India 
   \and Leibniz-Institut für Astrophysik Potsdam (AIP), An der Sternwarte 16, 14482 Potsdam, Germany
   }


 
  \abstract
  {The properties of the cool ($T\sim10^4$~K) gas in the circumgalactic medium (CGM) are closely linked to the physical mechanisms that create and maintain this multiphase medium.
  The cool CGM is thought to consist of discrete clouds, whose characteristic size is unknown. 
  Here we present a geometric and direct approach to constrain the coherence scale of these cool structures using stacked \mgii\ absorption lines measured against extended background galaxies and effectively point-like background quasars, whose sizes are a few kpc and \(\lesssim 0.01\)~pc, respectively.
  When the background-source size is smaller than the coherence scale of the foreground clouds, incomplete covering lowers the detection fraction and causes the median stacked absorption to differ from the mean.
  For stacked \mgii\ absorption against background galaxies, the mean and median equivalent width (EW) profiles are broadly consistent. 
  For stacked \mgii\ absorption against background quasars, by contrast, the median and mean EW profiles differ significantly, reaching a $\approx10\sigma$ difference beyond 100~kpc. 
  Furthermore, we find that the median and mean EW profiles are broadly consistent for large background galaxies (median half-light radius $\approx 6.6$~kpc), but differ at the $\approx3\sigma$ level for small background galaxies  ($\approx 1.5$~kpc).
  This suggests that the \mgii-bearing gas has an effective transverse coherence scale of order a few kiloparsecs, broadly consistent with the $\sim$2--7~kpc range probed by the background galaxies.
  Using a toy model in which the CGM is populated with discrete cool clouds, we show that the observed differences arise naturally from the combination of partial covering and beam averaging.
  Our results provide a new geometry-based measure of the small-scale structure of cool CGM gas.}
   \maketitle

\section{Introduction}
Galaxies evolve through a continuous exchange of mass, energy, and metals with their surrounding gaseous environment. 
The circumgalactic medium (CGM) acts as the interface between galaxies and the intergalactic medium (IGM), regulating the baryon cycle that governs galaxy growth and star formation \citep[e.g.][]{tumlinson17}. 
Within the CGM, gas and metals can be expelled from galaxies through stellar and AGN feedback, recycled through galactic fountains, or stripped from infalling satellites. 
At the same time, the CGM also hosts reservoirs of relatively metal-poor gas accreted from the IGM that can eventually cool and accrete onto galaxies, providing fuel for future star formation. 
Understanding the structure and dynamics of the CGM is therefore essential for building a complete picture of the galactic ecosystem.

Observationally, the CGM has been studied primarily through absorption-line spectroscopy against bright background quasars. 
This technique has provided powerful constraints on the spatial extent, kinematics, and chemical composition of circumgalactic gas over a wide range of redshifts \citep[e.g.][]{bahcall69,cowie95,schaye03,bouche12,nielsen13,werk14,borthakur15,lan18}. 
While such measurements are highly sensitive to low-density gas, they typically provide only a single, effectively pencil-beam sightline through each galaxy halo and therefore offer limited direct constraints on the transverse structure of the absorbing medium or the spatial coherence of individual absorbers.
Observations of the CGM in emission, particularly enabled by integral-field spectrographs, have begun to map the 2-dimensional distribution and kinematics of circumgalactic gas \citep[e.g.][]{wisotzki16,bacon21,guo23,guo24b,pessa24,pessa26}. 

Both theoretical models and numerical simulations suggest that the cool CGM phase traced by low-ionisation gas such as \mgii\ ($T\sim10^{4}$~K) is highly structured and clumpy within the hot, volume-filling gas halo.
A variety of physical processes can generate such multiphase structure \citep{faucher23}, including thermal instability and precipitation in hot halos \citep[e.g.][]{maller04,sharma12,voit15}, turbulent mixing and radiative cooling \citep[e.g.][]{kwak10,ji19,gronke20}, and the interaction of cool structures with hot halo flows, which can drive cloud disruption, fragmentation, and entrainment \citep[e.g.][]{armillotta17,gronke18}. Recent studies further show that radiative cooling in turbulent mixing layers can continuously regenerate cool gas, allowing cold material to survive and even grow within hot halos \citep[e.g.][]{gronke20,fielding20}. In addition, simulations of cloud fragmentation suggest that cooling instabilities can cause larger structures to break into numerous smaller cloudlets, producing a mist-like population of cold gas embedded within warmer material \citep{mccourt18}. Cosmological simulations of galaxy formation likewise predict that the CGM should contain a rich multiphase structure shaped by feedback, accretion, and mixing processes \citep[e.g.][]{oppenheimer18,nelson20}.
Quantitatively testing this picture requires observational constraints on the size distributions of cool clouds.

Direct imaging of individual cool clouds in the CGM is extremely challenging because of limited spatial resolution and the intrinsically low surface brightness of the gas. As a result, spatially resolved detections of small-scale cool structures are currently restricted to a few nearby systems, primarily in strong outflows, where individual filaments and cloud complexes are bright enough to be observed in emission. Well-studied examples include the starburst-driven wind of M82 \citep[e.g.][]{devine99,lehnert99,fisher25,lopez25,guo26} and a small number of similar local galaxies \citep[e.g.][]{heckman99}, but such cases are restricted to the local Universe and do not yet provide a general view of cool-gas structure throughout the broader CGM.

A number of approaches have been developed to constrain the characteristic sizes of CGM absorbers indirectly. One widely used method combines kinematic decomposition with photoionisation modelling of quasar absorption systems to infer the characteristic line-of-sight thickness of individual absorption components \citep[e.g.][]{schaye07,lan17,zahedy19,chen23}. 
Another approach exploits multiple background sightlines, such as quasar pairs or gravitational arcs, to probe the transverse coherence of absorbers \citep[e.g.][]{rauch99,rauch02,petitjean00,churchill03,ellison04,davis15,peroux18,augustin21,afruni23,lopez24,dutta24}. 
\citet{rubin18b} introduced a related approach by comparing the scatter in \mgii\ absorption measured toward bright background galaxies and background quasars, and showed that the difference implies a transverse coherence scale for cool CGM gas of at least 1.9~kpc.
More recently, constraints on very small-scale ionised structures have also been explored using radio propagation effects such as fast radio burst scintillation \citep{jow24}. 
These techniques probe a wide range of spatial scales but are often model dependent or rely on relatively rare observational configurations (e.g. multiple nearby quasars or gravitational arcs). 
A complementary, model-independent approach that can be applied to large statistical samples would therefore provide valuable additional constraints on the spatial structure of CGM absorbers.

In this paper, we introduce a geometry-based approach that exploits the difference between absorption measured against background quasars and against background galaxies.
Quasars are effectively point sources; 
most of their optical/UV continuum emission originates from gravitational energy released by gas accreting onto the black hole \citep[e.g.][]{antonucci93} and arises from a compact accretion disk with a characteristic size of \(\lesssim 0.01\)~pc \citep[e.g.][]{shakura73,morgan10}.
Background galaxies, in contrast, are extended sources with effective radii of several kiloparsecs.
These two classes of background sources probe the gas on very different transverse scales. 
A quasar traces a single pencil-beam sightline, whereas a background galaxy samples a bundle of neighbouring sightlines across its projected extent and therefore is more likely to intersect multiple foreground clouds.
The measured absorption against a galaxy is a column-density-- and surface-brightness--weighted average over many sightlines, and should depend on the effective size of the background beam and the size of the foreground clouds. 
In this way, comparing absorption against background sources of different sizes offers a direct geometric probe of the transverse coherence scale of cool CGM gas.

We apply this idea to \mgii\ absorption, a widely used tracer of cool ($T\sim10^{4}$~K) metal-enriched gas in the CGM at intermediate redshift \citep[e.g.][]{bouche12,lan18}. 
Our analysis is based on the MusE GAs FLOw and Wind (MEGAFLOW) survey \citep{schroetter16,schroetter19,schroetter21,schroetter24,zabl19,zabl20,zabl21,wendt21,freundlich21,langan23,cherrey24,cherrey25,bouche25}, which combines deep Multi-Unit Spectroscopic Explorer \citep[MUSE;][]{bacon10} integral-field spectroscopy with quasar absorption line measurements to identify large samples of foreground galaxies associated with intervening \mgii\ absorbers. 
The MEGAFLOW fields also contain numerous background galaxies that lie behind these foreground systems, enabling a direct comparison between absorption measured against quasars and against extended background galaxies within the same survey footprint.

This paper is organized as follows. In Sect.~\ref{sec_data}, we describe the MEGAFLOW survey data and the construction of our foreground–background galaxy sample, including the measurement of \mgii\ absorption in stacked spectra. In Sect.~\ref{sec_profiles}, we present the radial absorption profiles measured for background galaxies and compare them with those obtained from quasar sightlines. In Sect.~\ref{sec_model}, we introduce a toy model to interpret the observed differences in terms of beam averaging and the spatial distribution of CGM clouds. Finally, in Sect.~\ref{sec_discussion}, we discuss the implications of our results for the structure of cool gas in the CGM and summarize our main conclusions.
Throughout this paper, we adopt a $\Lambda$CDM cosmology with $\Omega_{\rm M}=0.307$, $\Omega_{\Lambda}=0.693$, and $H_{0}=67\,{\rm km\,s^{-1}\,Mpc^{-1}}$ \citep{plank15}. At the typical survey redshift, $z\sim1$, an angular scale of $1\arcsec$ corresponds to 8.23\,kpc. The magnitudes are given in the AB system. All distances are physical as opposed to co-moving. 

\section{Data}
\label{sec_data}

The MEGAFLOW survey is a VLT/MUSE + VLT/ Ultraviolet and Visual Echelle Spectrograph \citep[UVES;][]{dekker00} programme designed to connect cool-gas absorption in the circumgalactic medium (CGM) to the properties and kinematics of its host galaxies \citep{bouche25}. It targets 22 quasar fields selected from the SDSS \mgii\ absorber catalogue of \citet{zhu13}.
MEGAFLOW combines (i) 85\,hr optical integral-field spectroscopy with MUSE to identify and characterise galaxies in each quasar field, and (ii) 63\,hr high-resolution UVES spectroscopy to measure the absorption-line properties along the quasar sightlines. This survey strategy efficiently builds a statistical sample of absorber--galaxy associations within the $1\arcmin\times1\arcmin$ MUSE field of view. 

This work makes use of the MUSE data products from MEGAFLOW. 
The MUSE observations were obtained in visitor mode between 2014 and 2018 as part of Guaranteed Time Observations (GTO), using the Wide Field Mode (WFM) in both seeing-limited and adaptive-optics configurations \citep{bouche25}. 
The total exposure time per field ranges from $\sim$2 to 11\,hr. 
Data reduction followed the standard MUSE pipeline (v$\geq$2.4; \citealt{weilbacher20}) with additional post-processing steps developed for deep MUSE datasets, including inverse-variance combination and PCA-based sky-residual subtraction using ZAP \citep{soto16}. 
Galaxies were identified with a dual strategy that combines continuum-selected sources (from white-light images) and blind emission-line detections, enabling sensitivity to both passive systems and low-continuum line emitters. 
The resulting MUSE catalog contains $\approx$2400 sources, including 1403 galaxies with $\mathrm{ZCONF}>1$. 
$\mathrm{ZCONF}$ is the redshift-confidence flag defined by \citet{bacon23} and \citet{bouche25}. The criterion $\mathrm{ZCONF}>1$ selects sources with either a high-confidence single-line redshift, where the line shape is identifiable (e.g. a resolved \oii\ doublet or an asymmetric Ly$\alpha$ profile; $\mathrm{ZCONF}=2$), or a secure multi-line redshift ($\mathrm{ZCONF}=3$), while excluding low-confidence or unidentified redshift solutions.
Stellar masses ($M_\star$) are estimated via SED fitting with the \textsc{Coniecto} code \citep{zabl16}, assuming a \citet{chabrier03} stellar initial mass function; \mgii\ host galaxies in MEGAFLOW have a median stellar mass of $M_\star \simeq 10^{9.67}\,M_\odot$ \citep{bouche25}.
The star formation rates (SFR) are estimated using the \oii\ fluxes based on the empirical calibration of \citet{gilbank10,gilbank11}.

\subsection{Sample construction and spectral processing}
\label{subsec_sample_and_spectra}

\begin{figure*}
    \centering
    \includegraphics[width=1.9\columnwidth]{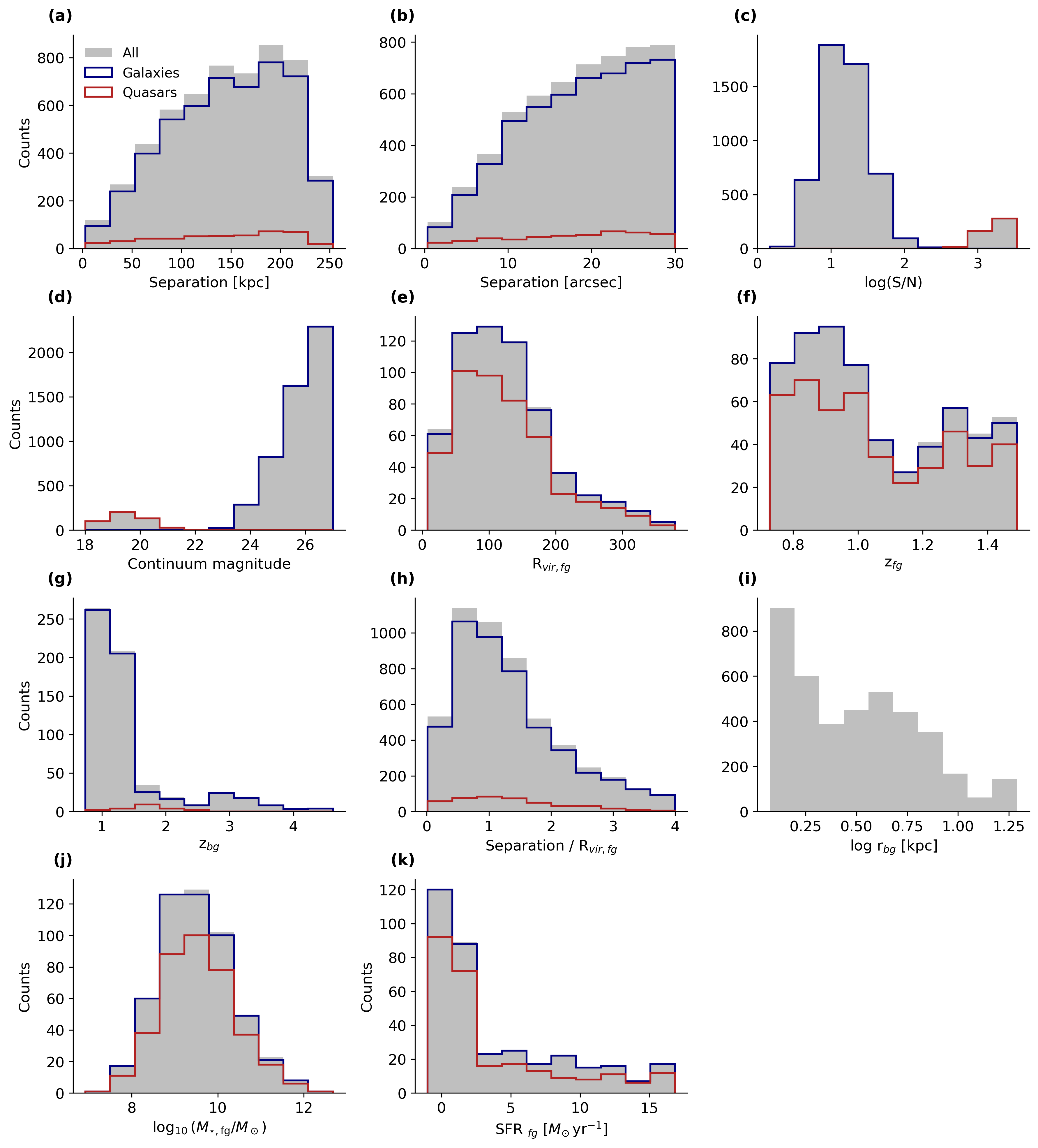}
    \caption{
    Distributions of key quantities for the final FG--BG pair catalogue used in this work. The grey histograms show the full sample, while the blue and red histograms indicate the background-galaxy and background-quasar subsamples, respectively. 
    (a) Projected separation in physical units ($b$, kpc). 
    (b) Projected separation in angular units (arcsec). 
    (c) Continuum signal-to-noise ratio of the extracted background spectrum. 
    (d) Continuum magnitude, measured in a 200\,\AA\ top-hat band centred at $\lambda_{\rm c}=2800(1+z_{\rm FG})$\,\AA. 
    (e) Foreground virial radius $R_{\rm vir,fg}$ (kpc). 
    (f) Foreground redshift $z_{\rm fg}$. 
    (g) Background redshift $z_{\rm bg}$. 
    (h) Separation normalised by the foreground virial radius, $b/R_{\rm vir,fg}$. 
    (i) Effective physical radius of the background galaxies, $r_{\rm bg}$ (kpc).
    (j) Stellar masses of the foreground galaxies.
    (k) SFRs of the foreground galaxies.
    }
    \label{fig_distributions}
\end{figure*}

Our analysis uses an extended version of the MEGAFLOW MUSE source catalogue, which includes necessary information such as galaxy ID, redshift, redshift confidence, broad-band photometry, $M_\star$, star formation rate (SFR) and derived structural parameters where available (e.g. size, Sérsic index, inclination, and position angle), following the description in Appendix~B of \citet{bouche25}.

We first construct a foreground (FG) galaxy catalogue by selecting objects with secure redshifts. In practice, we require a confident MUSE redshift solution ($\mathrm{ZCONF}>1$) and impose a redshift window such that \mgii\ falls within the MUSE wavelength coverage while [O\,\textsc{ii}] remains observable for robust systemic-redshift anchoring.
For each FG galaxy, we identify background (BG) sources within the MUSE field of view (a projected angular radius of $30\arcsec$) and require the BG redshift to exceed the FG redshift by $\Delta z>0.03$ to ensure a clear line-of-sight ordering. 
To prevent confusion between the foreground \mgii\ absorption region and intrinsic far-UV features of the background source, we further require the foreground \mgii\ wavelength to be redward of the background Ly$\alpha$ (with a small safety margin of 500~km/s).
The procedure yields a paired FG--BG catalogue comprising 10\,004 pairs. 

For each unique background galaxy or quasar, we extract a one-dimensional spectrum from the corresponding MEGAFLOW MUSE datacubes. We adopt an aperture size that depends on the apparent size of the background source: compact BG galaxies (radius $\leq 3$ pixels) and BG quasars use a $0.4\arcsec$ aperture, while more extended BG galaxies (radius $>3$ pixels) use a $1.0\arcsec$ aperture.
We mask intrinsic spectral features from the background sources, such as Ly$\alpha$, \ion{N}{V}, \ion{Si}{II} and its fluorescent emission lines, \ion{C}{II}, \ion{Si}{IV}, \ion{C}{IV}, \ion{He}{II}, \ion{O}{III}], and \ion{C}{III}], in the extracted spectra.
We model the continuum using a robust running-median filtering scheme with a width of 150~\AA, and generate both a continuum spectrum and a continuum-subtracted spectrum, similar to the approach adopted by previous studies \citep[e.g.][]{wisotzki16,guo24a}. 
From the continuum spectrum, we measure a ``pseudo-continuum'' magnitude at the expected observed wavelength of the FG \mgii\ feature using a top-hat bandpass of width 200\,\AA\ centred at $\lambda_{\rm c}=2800(1+z_{\rm FG})$\,\AA. 
We then normalize the continuum-subtracted profile by this local continuum level, producing a spectrum whose continuum is centred near zero.
We also apply a background-continuum cut, $m_{\rm cont}<27$\,mag, to the FG-BG pairs.
This threshold is chosen to maximise the number of usable sightlines while ensuring that the stacked spectra are not dominated by very low-S/N background sources. 
The resulting stacking sample contains 5503 FG--BG pairs, constructed from 610 distinct foreground galaxies and 597 distinct background sources. 

Finally, all spectra are shifted to the FG rest frame and resampled onto a common wavelength grid. 
In addition to the flux spectra, we produce corresponding continuum-normalised products on the same grid, which are used for stacking and equivalent-width measurements.

Figure~\ref{fig_distributions} summarises the properties of the final FG--BG pair catalogue used for stacking. 
Panels (a) and (b) show that the selected sightlines span projected separations from a few kpc out to $\simeq250$\,kpc (or $<30\arcsec$), providing broad radial coverage across the inner and outer CGM. 
Panel (h) presents the same separations normalised by the FG virial radii, indicating that the vast majority of pairs probe within a few virial radii. 
The virial radii are estimated using the stellar-to-halo mass relation from \citet{behroozi19}, as presented in \citet{cherrey25}.
Panels (c) and (d) quantify the background spectral quality. As expected, BG quasars are typically much brighter and have substantially higher continuum S/N than BG galaxies, while the galaxy-background sample occupies the faint, moderate-S/N regime most relevant for stacking. 
Panel (i) shows the distribution of effective background-galaxy radii, converted from the angular sizes to physical units at $z_{\rm fg}$, with a median of $\sim$2.1\,kpc. This provides the empirical basis for interpreting the galaxy-background stacks in terms of beam averaging across unresolved \mgii-absorbing structure.

\subsection{Stacking methodology}
\label{subsec_stacking}

\begin{figure*}[ht!]
    \centering
    \includegraphics[width=1.5\columnwidth]{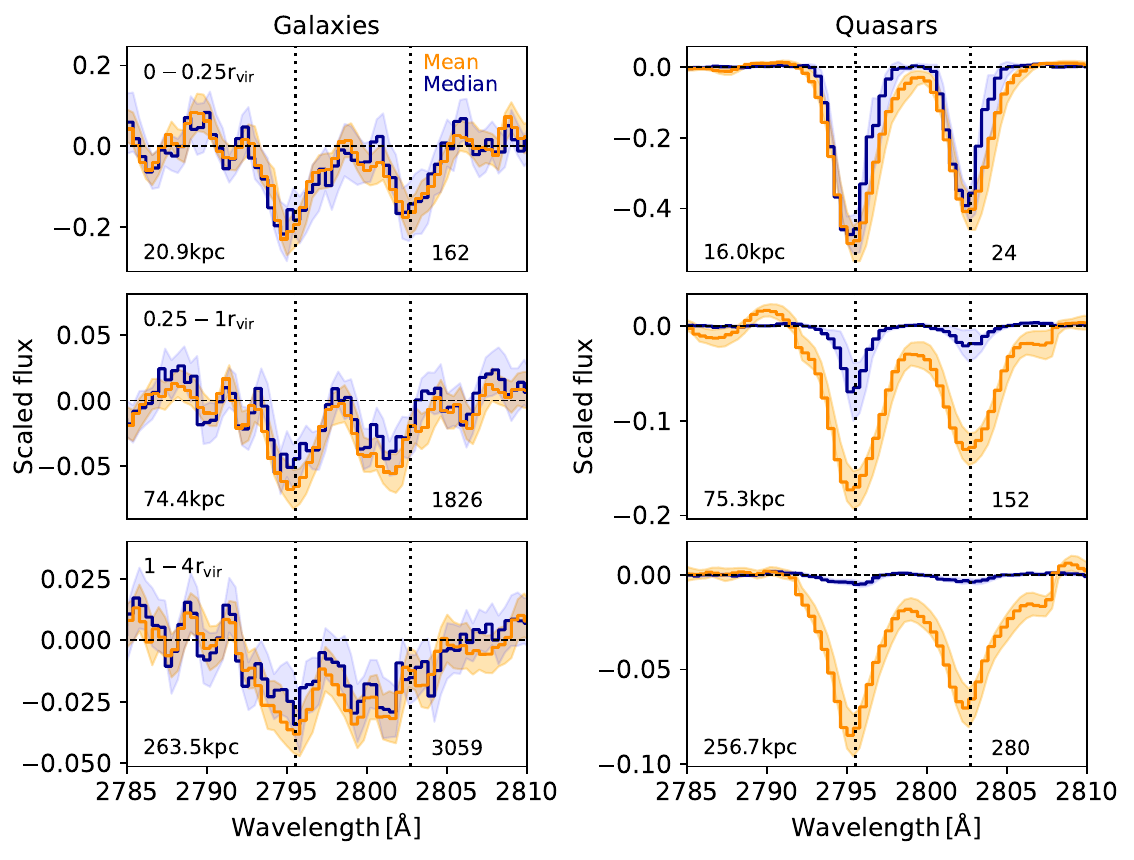}
    \caption{
    Stacked \mgii\ absorption profiles in the foreground rest frame for the final FG--BG pair catalogue. The three rows show bins of normalised separation $b/R_{\rm vir,fg}\in[0,0.25)$, $[0.25,1)$, and $[1,4)$. The left and right columns compare stacks constructed from background galaxies and background quasars, respectively. In each panel, the mean and median stacks are shown with different colours, and the shaded regions indicate the $1\sigma$ uncertainty envelopes estimated from bootstrap resampling. Vertical dotted lines mark the rest-frame wavelengths of Mg\,\textsc{ii} $\lambda2796$ and $\lambda2803$. The number of contributing sightlines and the median physical impact parameter of each bin are indicated in the panels. Note that the y-axis ranges differ between panels.
}
    \label{fig_stack_spec1}
\end{figure*}

We stack the \mgii\ absorption profiles using the rest-frame, continuum-subtracted and continuum-normalised spectra described in Section~\ref{subsec_sample_and_spectra}. Stacked profiles are constructed separately for background galaxies and background quasars.

We compute both mean and median stacks. 
To quantify how the average absorption varies with projected distance, we bin the pairs by separation normalised to the foreground virial radius, $b/R_{\rm vir,fg}$. 
We adopt three radial bins, $b/R_{\rm vir,fg}\in[0,0.25)$, $[0.25,1)$, and $[1,4)$, and compute galaxy-background and quasar-background stacks in each bin. 

Uncertainties on the stacked profiles are estimated via bootstrap resampling of the contributing spectra within each radial bin. Specifically, for each bin we draw the same number of spectra with replacement from the available sightlines, repeat the stacking, and iterate this procedure 100 times. We then take the standard deviation of the bootstrap-realisation stacks at each wavelength pixel as the $1\sigma$ uncertainty envelope. We compute bootstrap uncertainties separately for the mean stack and for the median stack, and propagate these uncertainties into measurements in Section~\ref{sec_profiles}. 

\section{\mgii\ absorption profiles} 
\label{sec_profiles}

\begin{figure}[ht!]
    \centering
    \includegraphics[width=1.\columnwidth]{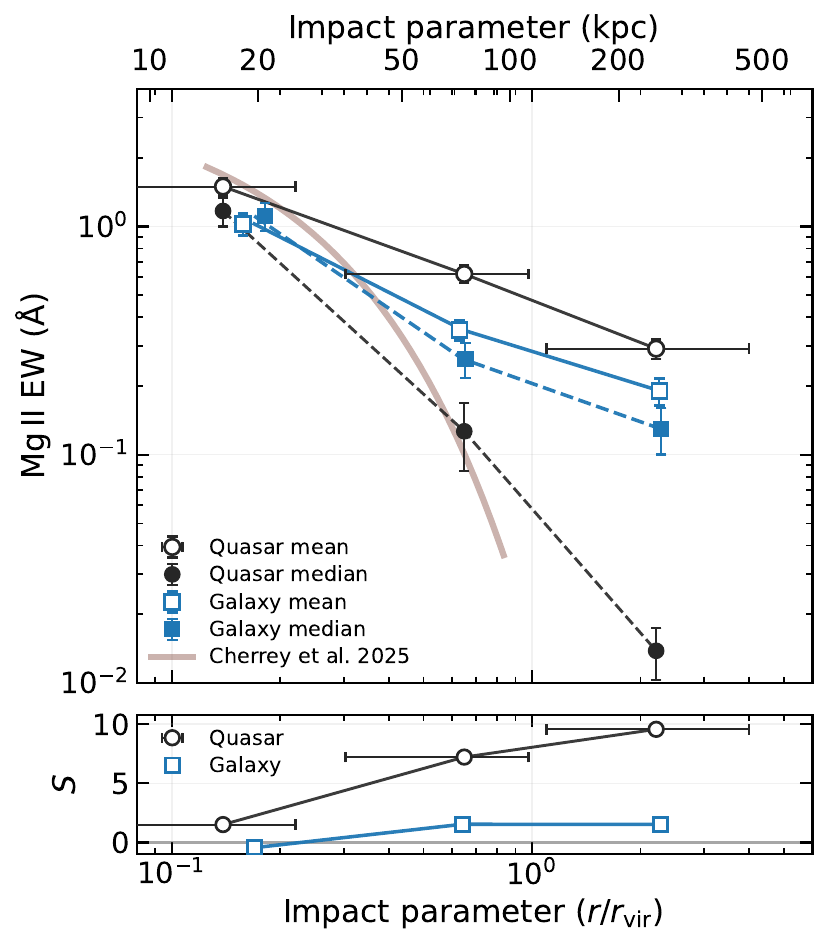}
    \caption{
    Radial dependence of the stacked \mgiiaw\ EW and the mean--median contrast. Top: EW measurements from the mean and median stacks constructed separately for background galaxies and background quasars in bins of normalised separation $r/R_{\rm vir,fg}$. The literature relation from \citet{cherrey25} for isolated galaxies measured against background quasars is over-plotted for comparison. 
    Bottom: significance of the mean--median contrast, $S=(EW^{\rm mean}-EW^{\rm median})/\sigma_{\Delta EW}$. The quasar stacks show a significant mean--median contrast at larger separations, whereas the background-galaxy stacks remain consistent between the two estimators.
    }
    \label{fig_EW1}
\end{figure}

\begin{figure*}[ht!]
    \centering
    \includegraphics[width=1.\textwidth]{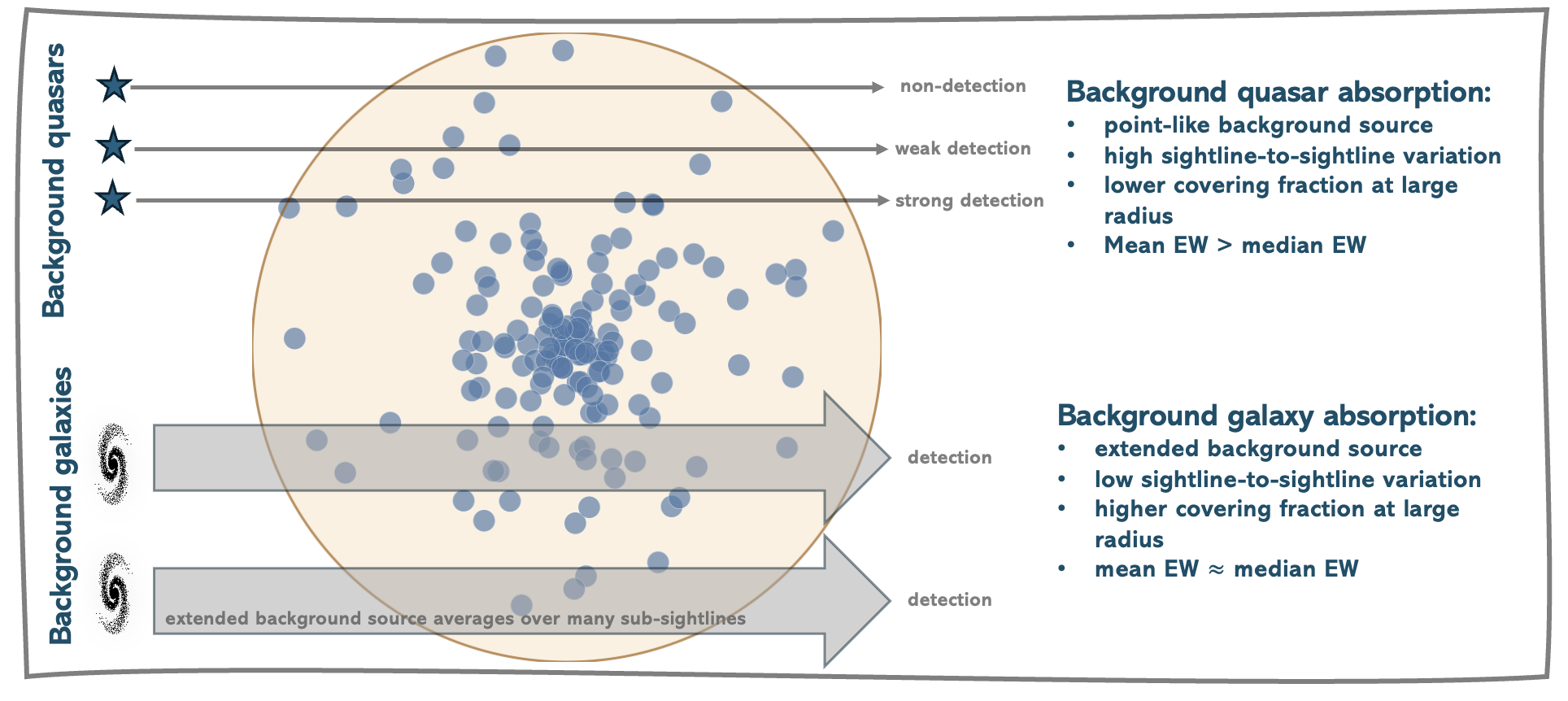}
    \caption{
    Schematic illustration of the physical picture explored in this paper. The cool \mgii-bearing gas is represented as a clumpy distribution of discrete structures within the halo of a foreground galaxy. For background quasars, each sightline probes a narrow pencil beam through the halo. For background galaxies, the source is extended, so the observed absorption averages over many sub-sightlines across the projected background light distribution. 
    }
    \label{fig_schematic_intro}
\end{figure*}

\begin{figure}[ht!]
    \centering
    \includegraphics[width=1.\columnwidth]{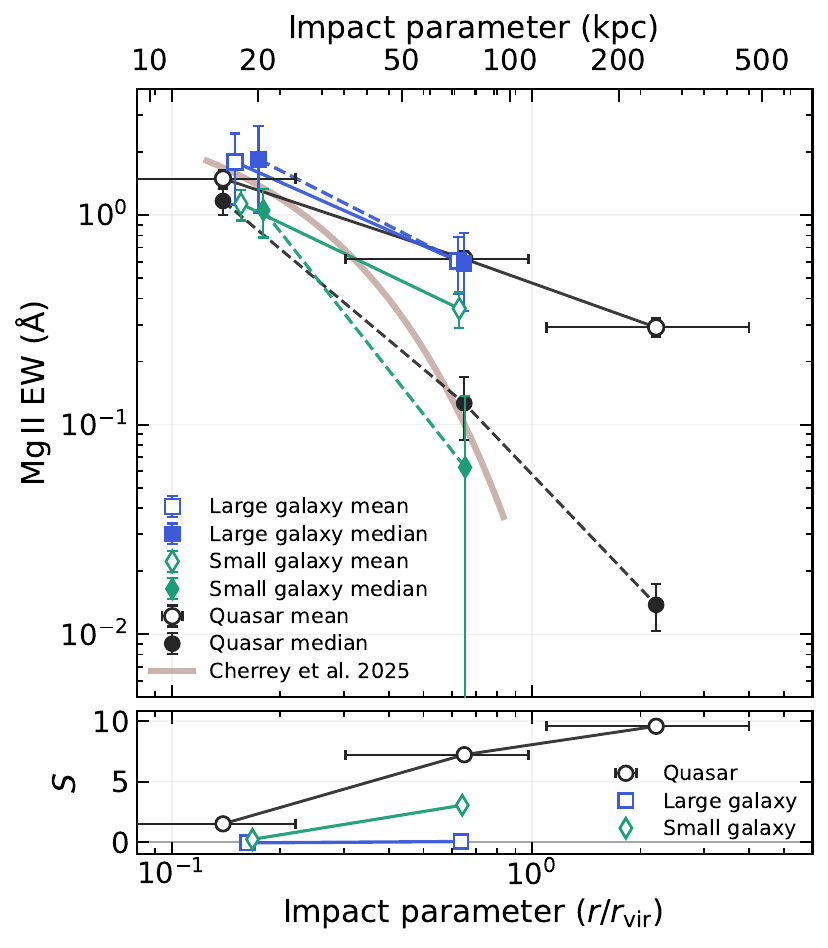}
    \caption{
    Same as Figure~\ref{fig_EW1}, but with the background-galaxy sample split into large ($r_{\rm bg}\approx 6.6$~kpc) and small ($r_{\rm bg}\approx 1.5$~kpc) sizes. 
    In the outermost radial bin, the S/N is insufficient for reliable EW constraints in the size-split galaxy stacks, and these measurements are therefore not shown.
    }
    \label{fig_EW_bins_size1}
\end{figure}


Figure~\ref{fig_stack_spec1} shows the stacked spectra in three bins of normalised separation, $b/R_{\rm vir,fg}\in[0,0.25)$, $[0.25,1)$, and $[1,4)$, comparing sightlines traced by background galaxies (left column) and background quasars (right column). In each panel, we show both the mean and median stacks, with shaded regions indicating the $1\sigma$ uncertainty envelopes. In all radial bins, the stacked spectra display the expected \mgii\ doublet at 2796 and 2803\,\AA\ in the foreground rest frame. 
The absorption weakens systematically with increasing $b/R_{\rm vir,fg}$. 
The innermost bin ($b/R_{\rm vir,fg}<0.25$) shows the strongest and most significant absorption, whereas the outermost bin ($1<b/R_{\rm vir,fg}<4$) shows substantially weaker absorption, approaching the noise level in the galaxy-background stacks. 
To quantify this radial trend, Figure~\ref{fig_EW1} presents the rest-frame equivalent width (EW) of the stacked \mgiiaw\ line as a function of projected separation.
The EWs are measured by summing the stacked absorption profile over a fixed rest-frame velocity window of $\pm250~{\rm km~s^{-1}}$ centred on the \mgii\ $\lambda2796$ transition. The corresponding measurements are listed in Table~\ref{tab:mean_median_ew}.

A key result of Figure~\ref{fig_EW1} is the contrasting behaviour of the mean and median EWs for the background-quasar and background-galaxy samples. 
Although the \mgii\ EW declines with radius for all stacks, the quasar mean EW decreases much more gradually than the quasar median EW. 
In contrast, the galaxy-background stacks show much closer agreement between the mean and median in all radial bins, and their radial trend is broadly similar to that of the quasar mean profile.
We quantify the significance of the difference between the mean and median stacks using
$S=(EW^{\rm mean}-EW^{\rm median})/\sigma_{\Delta EW}$, where
$\sigma_{\Delta EW}=[(\sigma_{EW,{\rm mean}})^2+
(\sigma_{EW,{\rm median}})^2]^{1/2}$, as is shown in the lower panel of Figure~\ref{fig_EW1}. The quasar stacks show a strong contrast, reaching $S\simeq10$ in the outermost bin, indicating that the mean EW is substantially larger than the median EW. In contrast, the background-galaxy stacks show much smaller contrasts, with $S\simeq1.5$ in the outermost bin, consistent with no significant mean--median difference.

These differences can naturally arise from a clumpy cool CGM with incomplete covering. Figure~\ref{fig_schematic_intro} schematically illustrates this picture, in which the cool \mgii-bearing CGM is distributed in discrete, clumpy structures within the halo of a foreground galaxy. A background quasar is effectively a point-like source, so each sightline probes only a narrow pencil beam through the halo, making the measured absorption highly sensitive to whether that sightline intersects a cool cloud. 
This naturally leads to strong sightline-to-sightline variation, a lower covering fraction at large radius, and a larger separation between the mean and median stacks. 
An extended background galaxy ($r_{\rm bg}\approx 2.1$~kpc in our sample), by contrast, averages over many sub-sightlines across the projected background light distribution. This beam-averaging effect suppresses the influence of rare clump interceptions, reduces the sightline-to-sightline variation, increases the probability of partial overlap with cool gas at a given radius, and yields a more stable absorption signal with closer mean--median agreement. Figure~\ref{fig_schematic_intro} is intended only as a schematic illustration of the main idea; a more quantitative physical model is presented in Section~\ref{sec_model}.

If the beam-averaging effect (Figure~\ref{fig_schematic_intro}) is indeed responsible for the observed differences between galaxy and quasar sightlines, then a similar effect should also appear within the background-galaxy sample itself: small and large background galaxies should not probe the foreground \mgii\ absorption in exactly the same way, especially if the background-source size is comparable to or smaller than the characteristic scale of the \mgii-bearing CGM structure. 
We examine how the stacked \mgii\ absorption depends on the effective size of the background galaxies; the corresponding EW radial profiles are shown in Figure~\ref{fig_EW_bins_size1}. 
The details of the subsampling and stacking procedure are presented in Appendix~\ref{app_size_dependence}. 
Splitting the sample into large ($r_{\rm bg}\approx 6.6$~kpc) and small ($r_{\rm bg}\approx 1.5$~kpc) background galaxies suggests a similar difference seen between background galaxies and quasars. 
Although the measurements are noisier, the small-background-galaxy stacks suggest a more noticeable mean--median difference ($\approx3.0\sigma$ in the intermediate bin), qualitatively following the same trend as seen for quasars, whereas the large-background-galaxy stacks remain more consistent between the two estimators ($\approx0.1\sigma$ in the intermediate bin), as is shown in the bottom panel of Figure~\ref{fig_EW_bins_size1}. 
This behaviour likely follows the interpretation that the observed \mgii\ absorption depends on how patchy cool gas is averaged over the projected extent of the background source. 
In this sense, the size dependence of the background-galaxy stacks suggests that the typical effective transverse coherence scale of the foreground cool gas is on scales of $\sim$~2-7~kpc, comparable to the range of background-galaxy sizes probed here.

For comparison, Figures~\ref{fig_EW1} and \ref{fig_EW_bins_size1} also show the literature $W_{2796}$--$b$ relation for isolated foreground galaxies and background quasars from \citet{cherrey25}, derived from the same MEGAFLOW dataset by fitting $\log_{10}(\mathrm{EW}) = A\,b + B$ to the individual detections and upper limits, where $b$ is the impact parameter. 
Our quasar median-stack measurements are consistent with this relation.
The galaxy-background mean and median EWs in Figure~\ref{fig_EW1} are also broadly consistent with previous background-galaxy stacking measurements \citep{bordoloi11,rubin18,santo26}, within the level of variation expected from the known dependence of the EW radial profile on host galaxy properties \citep[e.g.][]{chen10}.

The dependence of the stacked \mgii\ absorption on foreground stellar mass is examined in Appendix~\ref{app_mass_dependence}. We see possible mass-dependent variations in detectability and amplitude, particularly for the galaxy-background stacks, where absorption is significantly weaker (and in some bins not detected) for low-mass foreground galaxies. This behaviour may reflect a combination of intrinsically lower \mgii\ opacity around lower-mass haloes, and additional dilution effects at small impact parameters (e.g. \mgii\ emission partially filling the absorption). Importantly, however, the key empirical trends discussed in the main text remain unchanged: the quasar stacks continue to show a clear separation between the mean and median measurements, while the galaxy stacks show much closer agreement between the two estimators. 

We also test whether the different S/N properties of the quasar and galaxy spectra could drive the observed mean--median behaviour. Median stacks are more sensitive to noise than mean stacks \citep[e.g.][]{mishra24}, and adding noise can increase the apparent median EW when the intrinsic absorption is weak. In Appendix~\ref{app_noise}, we therefore add Gaussian noise to each individual quasar spectrum, matching the noise distribution of the background-galaxy spectra, and repeat the full stacking analysis. The noise-added quasar median EW declines more slowly with radius, but the quasar stacks still retain a clear mean--median contrast and do not reproduce the close mean--median agreement seen in the galaxy stacks. Thus, noise alone cannot explain the observed difference between the quasar and galaxy stacks.

\section{A toy model} 
\label{sec_model}

To interpret the systematic differences between the galaxy-background and quasar-background stacks, we construct a simple ``cloud-in-halo'' toy model. 
The model is broadly motivated by recent work \citep{hummels24,bisht25}, in which the cool CGM is described as a hierarchical population of kpc-scale cloud complexes, each composed of many smaller cloudlets, often on sub-kpc scales. 
In our case, however, the observations primarily constrain the transverse coherence scale relevant for extended background beams, so we adopt a simplified description in which the absorbing structures are treated as effective discrete kpc-scale clouds \citep[the ``cloud complexes'' in ][]{hummels24,bisht25}. 
The goal is to test whether a CGM populated by kpc-scale cool structures with incomplete covering can qualitatively reproduce (i) the stronger separation between the mean and median stacks seen for quasar (pencil-beam) sightlines and (ii) the more mean--median-consistent absorption measured against extended background galaxies. Additional sub-kpc structure within these effective clouds is not modelled explicitly here, but can be incorporated in future work if needed.

\subsection{Model setup: cloud in halo}
\label{subsec_model_setup}

\begin{figure}
    \centering
    \includegraphics[width=0.7\columnwidth]{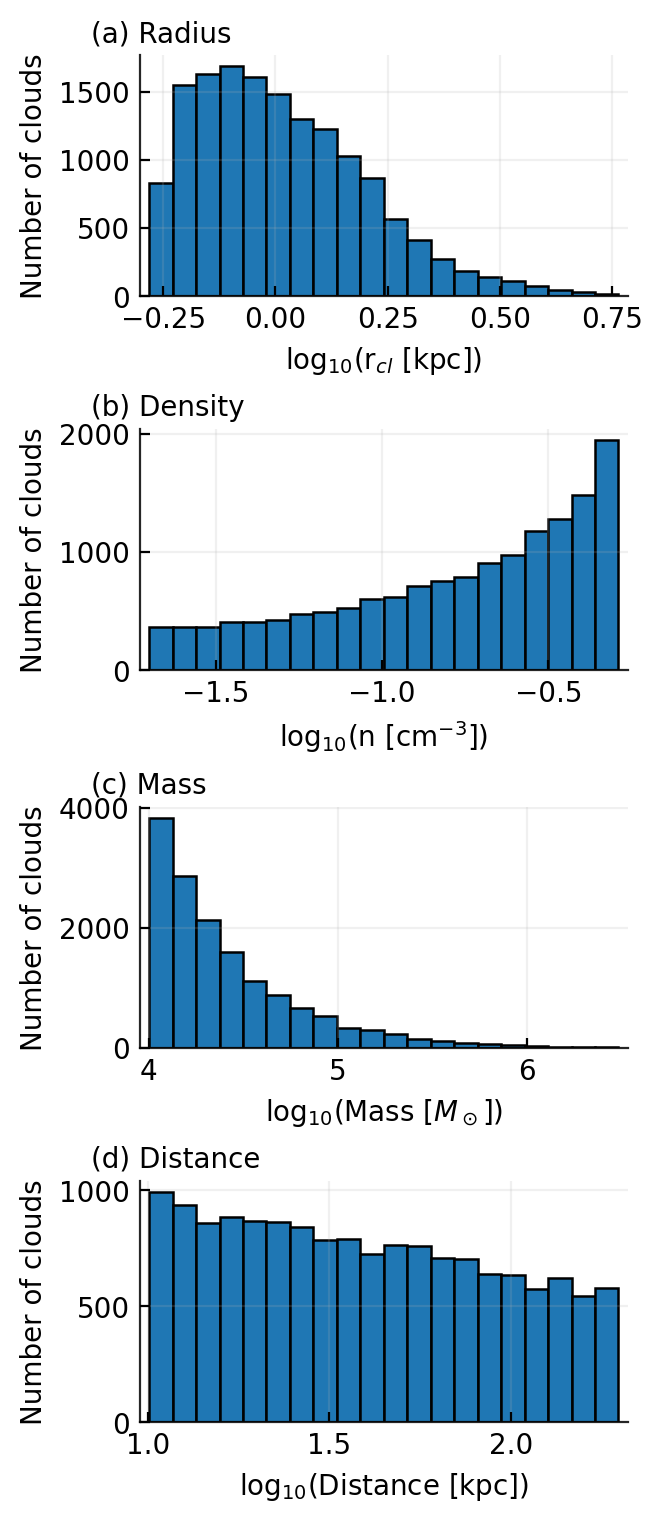}
    \caption{
    Distributions of physical properties for a representative realisation of the cool-cloud population by the toy model. Panels (a)–(d) show the distributions of cloud radius $r_{\rm cl}$, gas number density $n_{\rm cool}$, cloud mass $M_{\rm cl}$, and 3D galactocentric distance $r$.
    }
    \label{fig_cloud_distribution}
\end{figure}

\begin{figure*}
    \centering
    \includegraphics[width=0.49\textwidth]{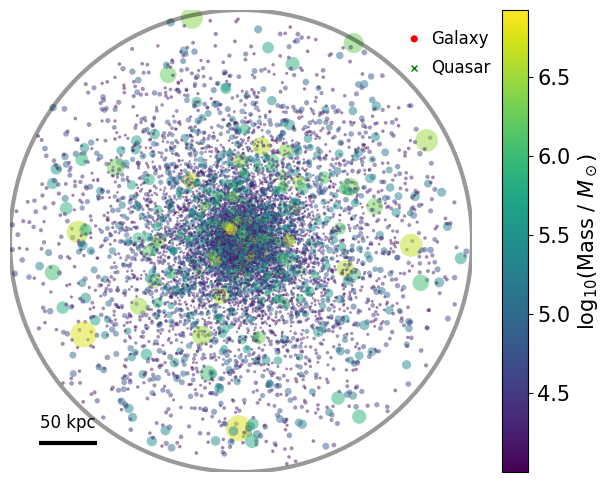}\hfill
    \includegraphics[width=0.49\textwidth]{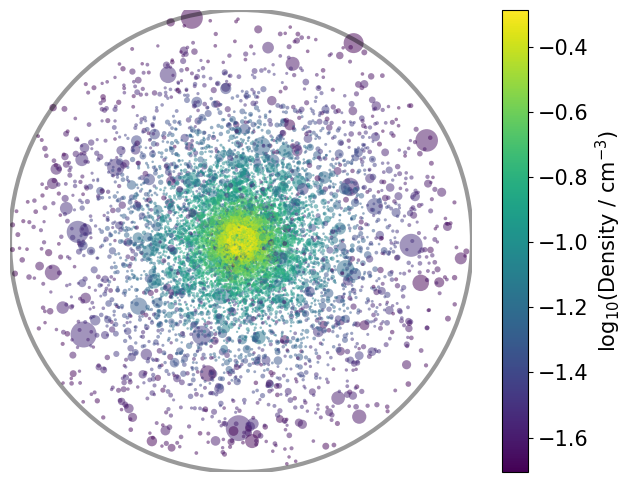}
    \caption{
    Representative projected realisation of the cloud population within the halo.
    Left: cloud positions, with marker size proportional to the cloud radius and colour indicating cloud mass.
    Right: same realisation, but coloured by the inferred cool-gas density $n_{\rm cool}$ from the pressure-equilibrium assumption (Equation~\ref{eq_nteq}).
    The grey circle marks the adopted outer halo boundary; the red circle and green cross denote the background galaxy and quasar sightlines used for the example projection.
    }
    \label{fig_mass_n_density}
\end{figure*}

We assume that cool \mgii-bearing clouds are embedded in a volume-filling hot halo. For the hot phase, we adopt the adiabatic hydrostatic halo model used by \citet{lan19}, who build on the multiphase cooling framework of \citet{maller04}. 
This model assumes a $M_{\rm h}=10^{12}\,{\rm M_\odot}$ NFW halo with concentration $c=10$. This profile is also broadly similar to that inferred for the Milky Way \citep[e.g.][]{gupta17}. 
We assume that the cool ($T_{\rm cool}\approx10^{4}\,$K) clouds are confined by the ambient hot halo, forming an approximate pressure balance,
\begin{equation}
n_{\rm hot}(r)\,T_{\rm hot}(r) \;=\; n_{\rm cool}(r)\,T_{\rm cool},
\label{eq_nteq}
\end{equation}
where $n$ is the hydrogen number density. 
This is motivated by the fact that a cloud can establish approximate thermal pressure balance with the surrounding hot medium on a sound-crossing timescale that is often shorter than its survival time \citep[e.g.][]{mo96,lan19}.
However, this pressure-confinement assumption should be regarded as a simplifying approximation. 
Given the stellar mass distribution (Figure~\ref{fig_distributions}) of our foreground sample, many systems may not host a stable, volume-filling hydrostatic hot halo. 
The adopted hot-halo profile is therefore not intended to provide a unique physical description of every foreground galaxy, but rather to set a plausible external pressure scale for estimating characteristic cloud sizes.

We populate the halo with an ensemble of cool clouds specified by a total cool-gas mass budget, $M_{\rm cool,tot}=10^{9}\,M_\odot$, given by the estimation of N.~Bouché et al. in preparation, and power-law distributions for the cloud masses and galactocentric radii.
In our fiducial realisation we adopt a bounded power-law cloud mass function, $\frac{{\rm d}N}{{\rm d}M}\propto M^{-\alpha_m}$,
with bounds $M_{\rm min}=10^{4}\,M_\odot$ and $M_{\rm max}=10^{7}\,M_\odot$.
We set power law slope $\alpha_m=2.0$, following the prediction of \citet{tan24} and \citet{augustin25}.
We draw clouds iteratively until the cumulative mass reaches $M_{\rm cool,tot}$, yielding $N_{\rm cl}=15131$ clouds in the fiducial run. 
Independently, each cloud is assigned a 3D galactocentric radius drawn from a bounded power-law radial distribution, $\frac{{\rm d}N}{{\rm d}r}\propto r^{-\alpha_r}$,
with $\alpha_r=1.2$ over $r_{\rm min}=10$ to $r_{\rm max}=200$\,kpc, and a random angular position, producing an isotropic spatial distribution.

Given each cloud's galactocentric distance, we compute its cool-gas volume density using Equation~(\ref{eq_nteq}) with $T_{\rm cool}=10^{4}\,$K. 
For a cloud of mass $M_{\rm cl}$ and density $n_{\rm cool}$, we then set its physical size by assuming a spherical geometry,
\begin{equation}
V_{\rm cl} \;=\; \frac{M_{\rm cl}}{\mu\,m_{\rm p}\,n_{\rm cool}\,\eta},\qquad
r_{\rm cl} \;=\; \left(\frac{3V_{\rm cl}}{4\pi}\right)^{1/3},
\label{eq_rcl}
\end{equation}
where $\mu m_{\rm p}$ is the mean mass per hydrogen atom, and we adopt $\mu\simeq 1.4$.
Here $\eta$ re-scales the effective volume of the cool \mgii-bearing gas cloud. 
We introduce $\eta$ for two related reasons. 
First, strict thermal pressure equilibrium may not hold in real CGM systems \citep{werk14,faerman23}. 
Observationally inferred cool-phase densities are often lower than expected from simple thermal pressure balance with the surrounding hot medium, possibly owing to additional non-thermal support and/or unresolved microphysics. 
Second, the observed absorption may arise from a mist of unresolved cloudlets embedded within a larger cold-gas complex.
Such a geometry naturally allows the cool phase to have a small volume filling factor while maintaining a large projected covering fraction \citep[e.g.][]{mccourt18,dutta24,bisht25}. 
Theoretical models and observational constraints suggest that cool CGM gas occupies only a small fraction of the halo volume, typically with a volume filling fraction of $\sim 10^{-4}$--$10^{-2}$ \citep[e.g.][]{mccourt18,vedantham19,dutta24}.
Here we adopt a fiducial value of $\eta=10^{-3}$. 
This value should be regarded as a first-order phenomenological approximation rather than a direct prediction of the pressure-equilibrium model. 
Future work could replace this simplified prescription by explicitly assembling individual cloudlets into cloud complexes, following approaches such as those of \citet{bisht25}, to provide a more physically resolved description of the absorbing structures.
Here our aim is not to reproduce the full ionisation, turbulent, magnetic, and pressure structure of the multiphase CGM, but rather to capture the first-order scalings that govern covering fraction, beam averaging, and the relation between cloud size and background-source size.

This toy model produces a wide dynamic range in cloud sizes: $r_{\rm cl}$ varies between $0.5$ and $6$~kpc, with a median value of $0.98$~kpc.
The resulting spatial distribution of clouds is illustrated in Figure~\ref{fig_mass_n_density}, which shows a representative projection of the cloud population within the model halo. 
This prescription leads to larger cloud sizes at larger radii for fixed cloud mass. 
This trend is consistent with the expectation that clouds embedded in the lower-pressure outer CGM are less compressed and therefore occupy larger effective volumes \citep[e.g.][]{lan19}. 
In the fiducial realisation, the median cloud size increases from $\approx0.75$~kpc at a galactocentric distance of 25~kpc to $\approx1.6$~kpc at 150~kpc.
This trend is qualitatively consistent with the simulations of \citet{augustin25}, who find that the CGM clumps have sphericalized radii from $\sim0.25$~kpc to several kpc, and that larger clumpy structures preferentially appear at larger galactocentric distances.
Figure~\ref{fig_cloud_distribution} summarises the resulting distributions of $r_{\rm cl}$, $n_{\rm cool}$, $M_{\rm cl}$, and $r$ for a representative realisation.

\subsection{Mock sightlines and synthetic absorption}
\label{subsec_mock_sightlines}

\begin{figure*}
    \centering
    \includegraphics[width=1.9\columnwidth]{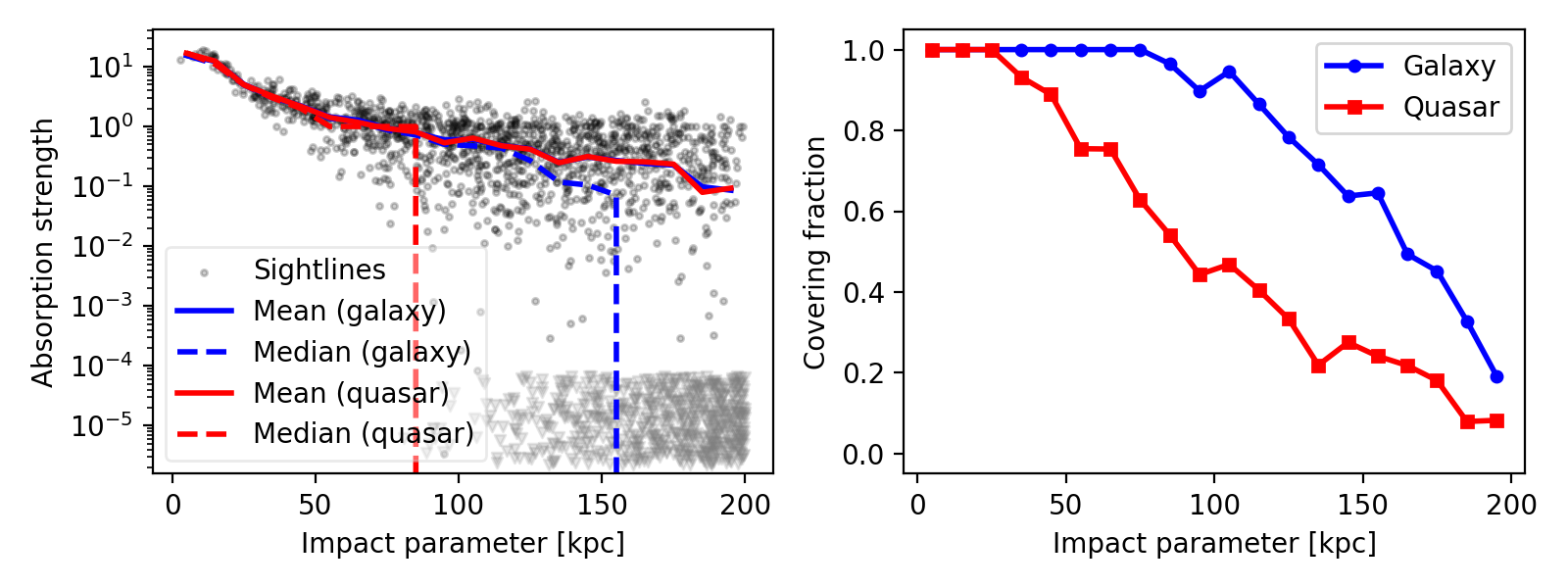}
    \caption{
    A representative realisation of the synthetic absorption calculation in the toy model.
    Left: Absorption strength as a function of impact parameter, defined as the fractional area of the background source covered by projected cool clouds. Grey points show individual galaxy-beam sightlines, while black downward triangles at the bottom mark non-detections, i.e. background-galaxy sightlines that do not intersect any foreground clouds. Solid curves show the mean profiles for extended background galaxies (blue) and pencil-beam quasars (red), while dashed curves show the corresponding median profiles.
    Right: covering fraction as a function of impact parameter, defined as the fraction of sightlines with non-zero cloud overlap. The different mean--median behaviour and the systematically higher galaxy-beam covering fraction arise naturally from partial covering and beam averaging across an extended background source.
    }
    \label{fig_model_profiles1}
\end{figure*}

Next, we forward-model mock sightlines. For each synthetic background position $(x,y)$ in the plane of the sky, we represent the background source as a circular beam of radius $r_{\rm bg}$ and compute its geometric overlap with every cloud in the same projection. 
We adopt $r_{\rm bg}=$2~kpc for background galaxies (motivated by the typical effective sizes in our pair catalogue) and a point-like beam for quasars ($10^{-4}$\,kpc in our fiducial realisation). 
We generate $N_{\rm sight}=2000$ random sightlines uniformly within the projected halo, and for each sightline compute an effective absorption strength as the summed fractional area covered by clouds,
$A \;\equiv\; \sum_{i}\,\frac{A_{{\rm overlap},i}}{\pi R_{\rm bk}^{2}}$,
where $A_{{\rm overlap},i}$ is the geometric overlap area between the source beam and the $i$th cloud. 
The resulting quantity $A$ is dimensionless and is used here as a proxy for the absorption equivalent width.
We also record a binary covering flag, ${\rm CF}=1$ if $A>0$ and ${\rm CF}=0$ otherwise, which traces the covering fraction of clouds as a function of impact parameter. All quantities are binned in projected impact parameter, and we compute the mean and median absorption strength, together with the covering fraction, in each radial bin.

Figure~\ref{fig_model_profiles1} shows one stochastic realisation of this procedure. The left panel plots the absorption strength $A$ versus impact parameter, including the distribution of individual galaxy-beam sightlines (grey points) and the corresponding mean and median radial profiles for galaxy (blue) and quasar (red) beams. The right panel shows the covering fraction profiles inferred from the same sightlines.

\begin{figure*}
    \centering
    \includegraphics[width=1.9\columnwidth]{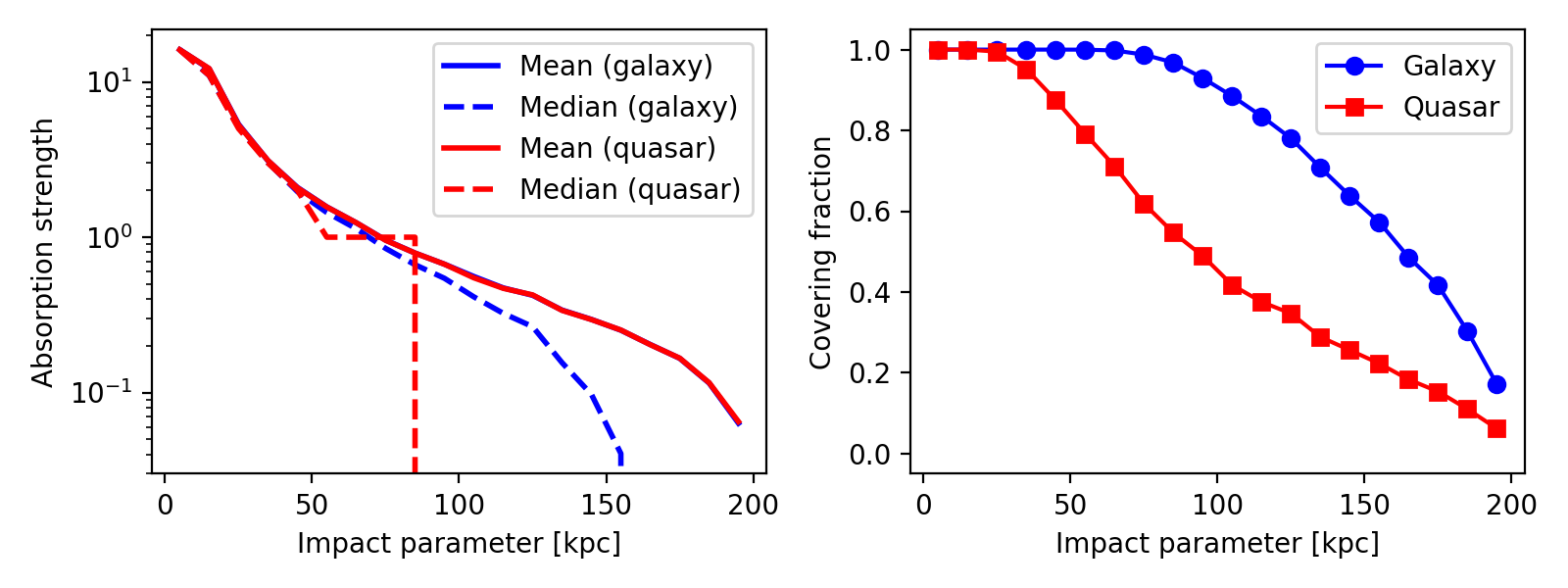}
    \caption{
    Monte-Carlo-averaged toy-model predictions. Left: absorption strength as a function of impact parameter for extended background-galaxy beams (blue) and pencil-beam quasar sightlines (red), showing both the mean (solid) and median (dashed) estimators. Right: corresponding covering fraction profiles for galaxy (blue) and quasar (red) beams, defined as the fraction of sightlines with non-zero cloud overlap. 
    }
    \label{fig_model_profiles2}
\end{figure*}

To reduce sensitivity to any single random draw, we repeat the full calculation $N_{\rm MC}=200$ times and average the resulting radial profiles. In each Monte-Carlo realisation we generate a new cloud population by sampling the given distributions, and draw $N_{\rm sight}$ random background absorptions.
Figure~\ref{fig_model_profiles2} shows the Monte-Carlo-averaged model predictions. The left panel presents the radial profiles of the synthetic absorption strength, while the right panel shows the corresponding covering fraction profiles. 
This cloud-in-halo model provides a plausible physical explanation for the statistical behaviour in observations.
Within $\approx$25\,kpc, the model robustly reproduces the near-unity covering fraction commonly inferred from quasar absorption-line statistics \citep[e.g.][]{cherrey25}. 
At larger radii, the quasar (pencil-beam) sightlines exhibit a substantially lower covering fraction at fixed impact parameter, so the median absorption strength declines rapidly, dropping to 0 at $\approx$80~kpc. 
In contrast, the extended galaxy beam maintains a higher covering fraction to larger radii, because its larger projected area increases the probability of intersecting at least one cloud. 
As a result, the model naturally produces a smaller separation between mean and median for galaxies than for quasars, qualitatively consistent with the behaviour seen in the stacks (Figure~\ref{fig_EW1}).

Overall, repeating and averaging over many stochastic realisations demonstrates that the distinctive mean--median contrast between quasar and galaxy backgrounds is not an artefact of a particular cloud configuration. The toy model thus provides a physical link between the small-scale cloud statistics (sizes, covering fraction) and the key observational trends in Figures~\ref{fig_stack_spec1} and \ref{fig_EW1}: quasar (pencil-beam) sightlines naturally exhibit larger sightline-to-sightline variance and a stronger mean--median separation, while extended background sources yield systematically more stable absorption through beam averaging.

This toy model naturally produces an asymmetric distribution of foreground absorption strengths. The variation of this distribution with galactocentric distance leads to the radial trends in the mean--median difference. The model also produces the radial covering-fraction behaviour, from the near-unity covering fraction in the inner CGM to the decreasing covering fraction in the outer CGM, as observed in \mgii\ absorption surveys \citep[e.g.][]{cherrey25}.
We emphasize, however, that the observed mean--median contrast is not unique to this toy model. In general, any asymmetric distribution of foreground absorption strengths will produce different mean and median EWs. The toy model therefore provides one physically motivated explanation for the observed contrast, but does not uniquely determine its physical origin.

We note that modelling absorption against extended background galaxies in a fully realistic way introduces additional complexity beyond the simplified treatment adopted here.
In the toy model, the absorption strength is defined only by the geometric overlap between the background beam and the projected cool clouds, assuming that the overlapping regions are optically thick. 
This approximation is motivated by the stacked \mgii\ profiles (Figure~\ref{fig_stack_spec1}), for which the \mgii\ $\lambda2796/\lambda2803$ doublet ratio is close to unity, indicating that the absorption is dominated by optically thick gas. 
Nevertheless, assigning an \mgii\ column density or optical depth to each toy cloud would require additional assumptions about the metallicity, ionisation state, internal density structure, velocity field, and unresolved substructure of the clouds. 
We therefore do not attempt to predict the \mgii\ EW or saturation fraction from the toy model; instead, we use it only to capture the first-order trends.

Several other effects may affect the quasar- and galaxy-background mean EWs. First, absorption against an extended background galaxy is a surface-brightness--weighted average over an uneven background light distribution, rather than a simple geometric average. The low-surface-brightness outskirts of a background galaxy span a larger projected area and may therefore have a higher probability of intersecting foreground clouds, while uncovered or only weakly covered bright central regions contribute a disproportionate fraction of the continuum flux and can dilute the absorption trough. This effect tends to reduce the observed background-galaxy EW relative to a pencil-beam quasar sightline. The impact of this non-uniform background coverage will be discussed further in K.~Gohlke et al.\ in preparation. 
Second, when the \mgii\ doublet is saturated, the EW depends not only on column density or covering fraction, but also strongly on the velocity spread of the absorbing gas along the sightline.
If several saturated clouds overlap both in projection and in line-of-sight velocity, they do not produce more absorption than a single saturated cloud covering the same part of the background beam. However, if the clouds are separated in line-of-sight velocity, they can increase the EW by broadening the absorption profile.
An extended background galaxy may therefore have a higher probability of intersecting multiple clouds with different line-of-sight velocities, whereas a quasar pencil beam is more likely to intersect only one or a few clouds. This effect can increase the galaxy-background EW relative to the quasar-background EW.
Thus, both the background light distribution and the foreground velocity distribution can affect the observed EW. These effects can act in different directions and are not included in our simplified model. We therefore use the toy model only to interpret the qualitative mean--median behaviour, rather than the absolute EW normalisation or the detailed offset between the quasar- and galaxy-background mean profiles. A more complete model would require coupling cloud covering statistics to cloud kinematics and to the resolved surface-brightness profiles of the background galaxies.

\subsection{Model parameter variation}
\label{subsec_parameter_variation}

\begin{figure*}
    \centering

    \begin{subfigure}[t]{0.26\textwidth}
        \centering
        \includegraphics[width=\textwidth]{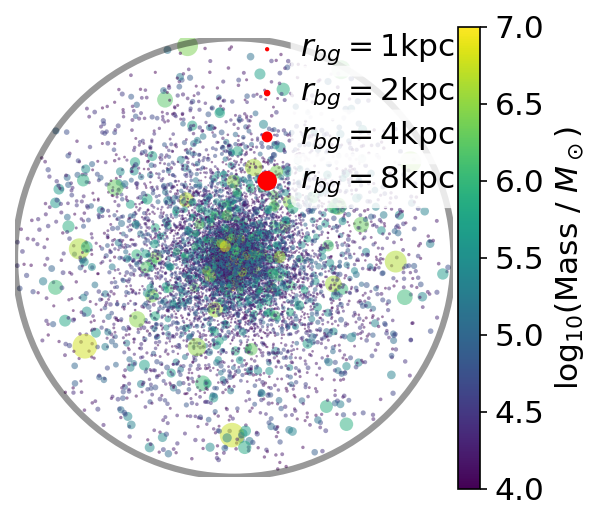}
    \end{subfigure}\hfill
    \begin{subfigure}[t]{0.69\textwidth}
        \centering
        \includegraphics[width=\textwidth]{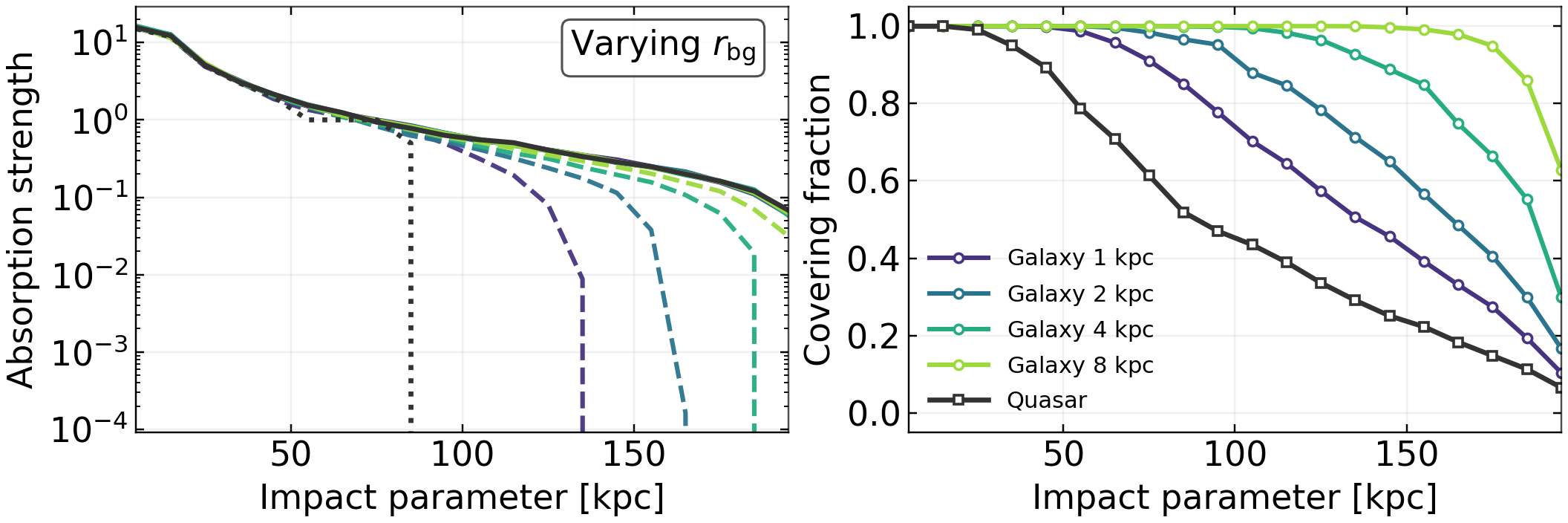}
    \end{subfigure}

    \vspace{0.5em}
    \rule{\textwidth}{0.3pt}
    \vspace{0.7em}

    \begin{subfigure}[t]{0.26\textwidth}
        \centering
        \includegraphics[width=\textwidth]{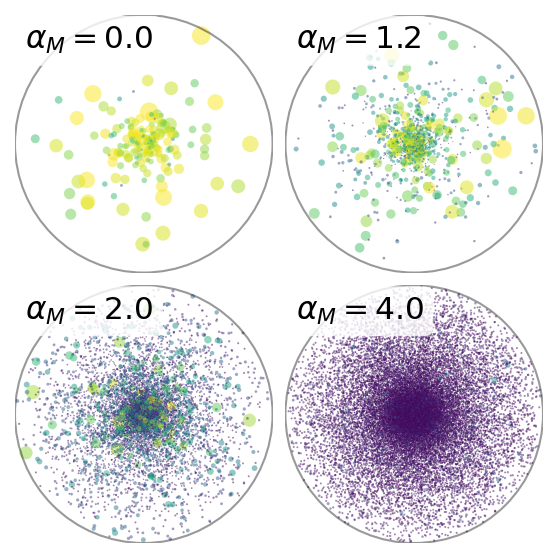}
    \end{subfigure}\hfill
    \begin{subfigure}[t]{0.69\textwidth}
        \centering
        \includegraphics[width=\textwidth]{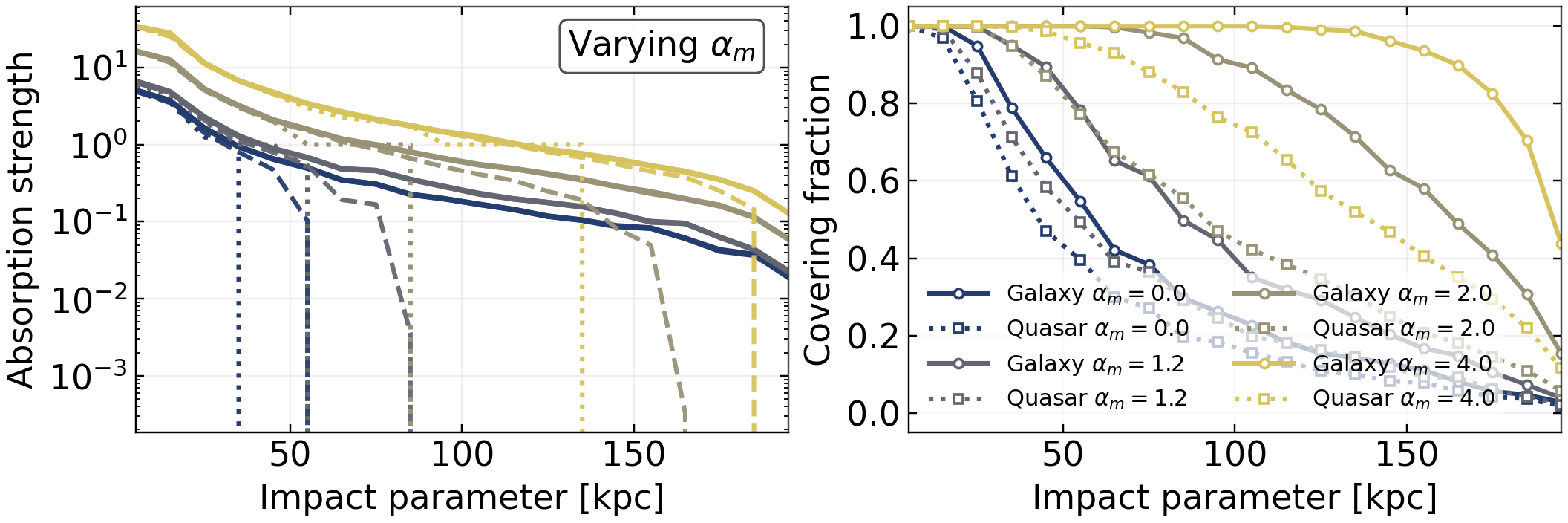}
    \end{subfigure}

    \vspace{0.5em}
    \rule{\textwidth}{0.3pt}
    \vspace{0.7em}

    \begin{subfigure}[t]{0.26\textwidth}
        \centering
        \includegraphics[width=\textwidth]{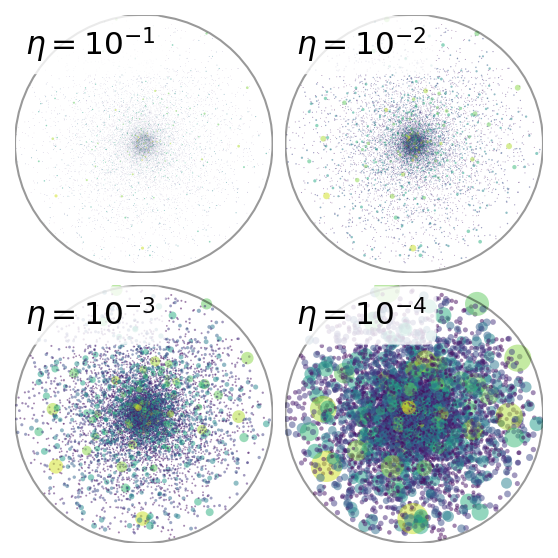}
    \end{subfigure}\hfill
    \begin{subfigure}[t]{0.69\textwidth}
        \centering
        \includegraphics[width=\textwidth]{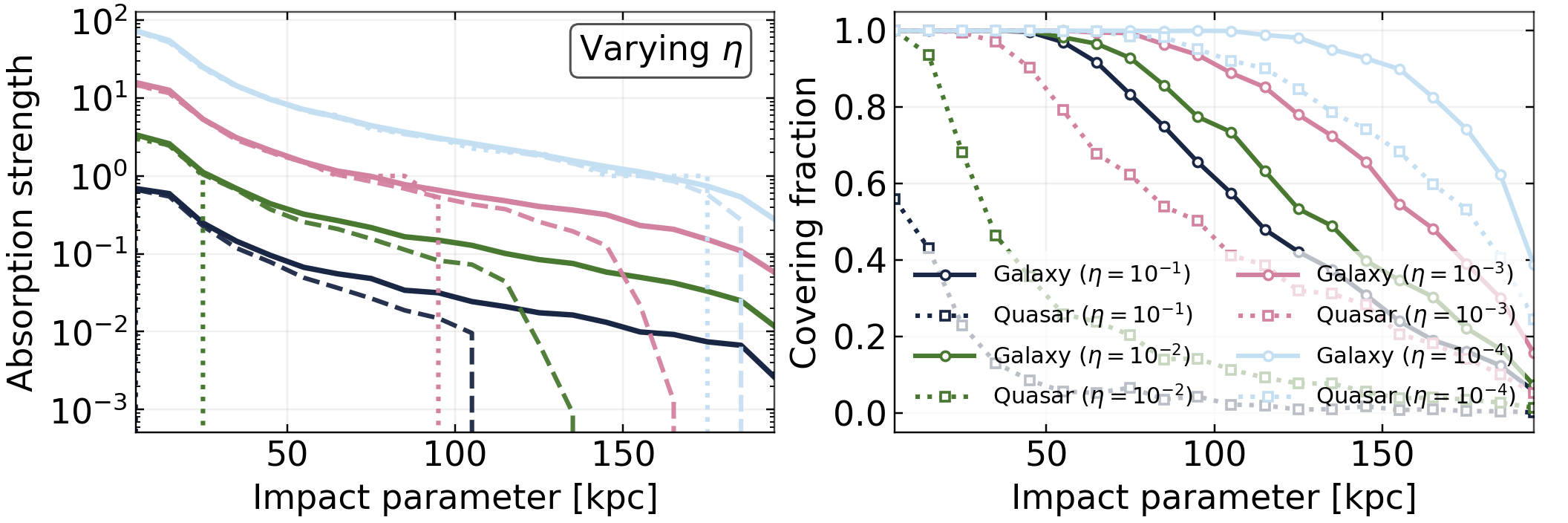}
    \end{subfigure}

    \caption{Predicted observational signatures of varying key parameters in the toy model. 
    In each row, the left panel shows the corresponding cloud mass distribution, while the middle and right panels present the resulting radial profiles of synthetic absorption strength and covering fraction, respectively. 
    The top row shows the effect of varying the effective radius of the background-galaxy beam. 
    The middle row shows the effect of varying the slope of the cloud mass function, $\alpha_{\rm m}$, with larger $\alpha_{\rm m}$ producing more numerous small clouds. 
    The bottom row shows the effect of varying the parameter $\eta$, which changes the cloud sizes while leaving the cloud number and spatial distribution unchanged. 
    All parameters other than the one being varied are fixed to the fiducial values. 
    In the middle column, the solid lines denote the mean profiles for quasar and galaxy backgrounds, which are nearly identical in this simplified model, while the dotted and dashed lines denote the median profiles for quasar and galaxy backgrounds, respectively.
    }
    \label{fig_model_profiles_variation}
\end{figure*}

The toy model described above contains several empirical parameters that regulate the cloud population in the halo, including the slope of the cloud mass function ($\alpha_{\rm m}$), the slope of the radial cloud distribution ($\alpha_{\rm r}$), the allowed cloud mass range ( $M_{\rm min}$ and $M_{\rm max}$), the total cool-gas mass budget, and $\eta$ that controls the size of the cool clouds. 
The adopted hot-halo profile introduces further model choices, including the halo mass, concentration, cooling history, and hot-gas density normalisation \citep{mo96}.
In addition, the effective size of the background beam sets the scale over which absorption is averaged. 
Although the current stacked data do not yet provide sufficient S/N to constrain these parameters precisely, we show here that the model already predicts distinct and observable signatures as the parameters are varied. However, these signatures are not unique, and different combinations of parameters can produce similar observable trends.

Increasing the background-beam size systematically raises the covering fraction at a given impact parameter, because a larger beam has a higher probability of intersecting at least part of the cloud population. Consequently, the median absorption strength remains elevated to larger distances for larger background sources compared to smaller ones. 
As $r_{\rm bg}$ decreases, the median absorption profile for galaxy backgrounds diverges more strongly from the mean and approaches the quasar median profile. This is qualitatively consistent with what we find in the size-split stacks presented in Figure~\ref{fig_EW_bins_size1} and Appendix~\ref{app_size_dependence}. 
For example, at an impact parameter of approximately $150$~kpc, the model with $r_{\rm bg}\approx2$~kpc produces relatively similar mean and median absorption strengths, whereas the model with $r_{\rm bg}\approx1$~kpc retains a clear mean--median separation. At this distance, the median cloud size in the fiducial realisation is approximately $1.6$~kpc. Similarly, in our observations, the small-background-galaxy stack ($r_{\rm bg}\approx1.5$~kpc) shows a tentative mean--median separation. This qualitative agreement suggests that the effective transverse coherence scale of the foreground cool gas cannot be substantially smaller than the small-background-galaxy beam size and is therefore likely of order kiloparsecs.

The middle panels of Figure~\ref{fig_model_profiles_variation} illustrate the impact of varying the cloud mass-function slope $\alpha_{\rm m}$ while keeping the background-galaxy beam fixed. 
The parameter $\alpha_{\rm m}$ controls the relative abundance of low-mass versus high-mass clouds: shallower mass functions (smaller $\alpha_{\rm m}$) favour a larger fraction of massive large clouds and therefore a smaller total number of clouds, whereas a larger $\alpha_{\rm m}$ populates the halo with many more low-mass small clouds. 
At fixed total cool-gas mass, $M_{\rm cool,tot}=10^{9}\,M_\odot$, increasing $\alpha_{\rm m}$ from 0 to 4 raises the number of clouds by a factor of $\approx 323$, from 207 to 66\,962. 
This increase in cloud number density boosts both the covering fraction and the absorption strength for quasar and galaxy beams alike. 
The separation between the mean and median profiles becomes progressively smaller as $\alpha_{\rm m}$ increases. Physically, once the halo is populated by a very large number of clouds, the absorber distribution becomes less patchy on the scale of the background beam, so individual sightlines are less dominated by rare intersections and more representative of the ensemble average.

The bottom panels of Figure~\ref{fig_model_profiles_variation} show the effect of varying $\eta$. This parameter serves as a convenient proxy for changing the cloud size while keeping the cloud number and spatial distribution fixed. Increasing $\eta$ produces smaller clouds. 
We vary $\eta$ from $10^{-1}$ to $10^{-4}$, with $\eta=10^{-3}$ adopted as the fiducial value. Relative to the fiducial realisation, the cloud radius increases by a factor of $\simeq2.15$ for $\eta=10^{-4}$ and decreases by factors of $\simeq0.46$ and $\simeq0.22$ for $\eta=10^{-2}$ and $10^{-1}$, respectively. Larger clouds have larger geometric cross-sections and therefore produce higher covering fractions and stronger absorption at fixed impact parameter for both quasar and galaxy beams.

Together, these experiments demonstrate that the toy model produces distinct observational signatures linked to both the intrinsic cloud population and the effective size of the background beam. However, these signatures are degenerate across several model parameters, such that different parameter choices can give rise to similar observable trends. Moreover, given the limited sample size and S/N of the current dataset, the simplified nature of the toy model and the strong assumptions underlying it, we do not attempt a formal fit to the data. 
Future observations with larger foreground--background galaxy samples and improved S/N should enable tighter constraints through joint modelling of the absorption-strength profiles and covering-fraction statistics.

Another notable trend is that, regardless of the parameter variation explored here, the quasar median absorption strength in the toy model drops sharply to zero beyond a characteristic impact parameter, once the covering fraction falls below 50\%. By contrast, the observed quasar median declines more gradually with radius (Figure~\ref{fig_EW1}), although it remains steeper than the quasar mean. This suggests that the real stacked signal is not well represented by a single toy-model realisation, but instead likely arises from a mixture of foreground haloes spanning a range of physical conditions and cloud configurations. Moreover, the present model assumes that all absorption arises within the foreground halo and neglects the two-halo contribution from neighbouring structures along the line of sight. This two-halo term can become important at large distances \citep[e.g.][]{ho20}, and its omission likely contributes to the overly sharp cutoff predicted by the toy model. Future datasets with substantially larger samples should make it possible to define subsamples of foreground galaxies by their physical properties and test whether different observational trends correspond to different regions of model parameter space.

\section{Discussion}
\label{sec_discussion}

One method that has been used to constrain cloud sizes is to infer the characteristic absorber thickness along the line of sight from quasar absorption, adopting $l_{\rm cl}\sim N_{\rm HI}/[f_{\rm HI}\,n_{\rm H}]$,
where $n_{\rm H}$ is the hydrogen volume density and $f_{\rm HI}$ is the neutral fraction. 
In practice, metal ions are often used instead of \hi. 
These quantities are typically constrained by matching observed ionic column densities with photoionisation models under an assumed radiation field \citep[e.g.][]{schaye07, lan17,chen23}. 
While powerful, these inferences can depend sensitively on the adopted photoionisation framework and on the fact that a single ``kinematic component'' in velocity space may blend multiple unresolved sub-components and/or gas phases with different densities and ionisation states \citep{marra24}. 
The characteristic scales for metal lines inferred in this way are often well below kpc \citep[e.g.][]{schaye07,chen23}. This does not contradict our findings: the absorption-inferred thickness may trace dense substructures embedded within a larger complex that sets the opacity and covering in projection. Our present stacks cannot resolve sub-kpc cloudlets, and the cloud size constrained by beam averaging should be interpreted as an effective transverse coherence/covering scale of \mgii\ absorption across an extended background source, rather than the physical thickness of a single density peak along the line of sight.

Another class of constraints comes from direct transverse coherence measurements using multiple nearby sightlines, such as quasar pairs, lensed quasars, and extended lensed arcs \cite[e.g.][]{rauch02,davis15,peroux18,augustin21,lopez24,dutta24,shaban25}. Such systems provide an empirical handle on the spatial coherence of gas, with inferred coherence scales ranging from $\lesssim$1~kpc to $\approx$10\,kpc depending on ion, system, and definition of ``coherence'' \citep[e.g.][]{augustin21,lopez24}. 
Studies targeting \mgii\ often find kpc-scale coherence in at least a subset of absorbers, together with substantial inhomogeneity on similar scales \citep{peroux18}. A recent study of three giant gravitational arcs further found that the coherence scale of \mgii-bearing gas lies between 1.4 and 7.8~kpc \citep{afruni23}.
Our galaxy-background stacks probe a similar regime of effective transverse averaging, but do so statistically across a large sample. In this way, our method exchanges the fine spatial resolution offered by rare lensed systems for statistical power, enabling population-level constraints on absorber coherence and covering fractions.

\citet{rubin18b} constrained the coherence scale of cool CGM gas by comparing the scatter in the \mgii\ EW--impact parameter relation measured with background quasars and extended background galaxies. 
In their framework, the CGM is effectively tiled by \mgii-bearing structures (thus the covering fraction is always unity), and the observed scatter arises from spatial variations in absorption strength across the halo. 
Based on an empirical EW--$b$ relation from individual background-quasar sightlines, they simulated how the scatter relative to this relation would change when averaged over the larger beam of a background galaxy. 
If the absorber structure fluctuates on scales much smaller than the projected size of the background galaxy, beam averaging should strongly suppress the scatter in the EW--$b$ relation. 
By comparing these predictions with the observed distribution toward background galaxies, they argued that the measured large scatter in background galaxy absorption cannot be produced by small-scale fluctuations, and inferred a characteristic \mgii\ coherence length of $\gtrsim 1.9$~kpc. 
Our results are broadly consistent, but our observational strategy is different. 
Rather than relying on an empirical EW--$b$ relation and robust background galaxy absorption detections, we compare the radial behaviour of the mean and median stacked absorption profiles for background galaxies and quasars, which sets constraints on both the absorption strength and covering fraction. 

Our results are broadly consistent with the emerging theoretical picture that the cool CGM is not a smooth phase but a clumpy, multiphase medium shaped by fragmentation, turbulent mixing, and interactions between cold structures and a hotter ambient halo \citep[e.g.][]{mccourt18,gronke18,gronke20,fielding20,nelson20,bisht25,ramesh24,hummels24}. High-resolution simulations show that radiative mixing layers and turbulent interfaces can continuously generate and entrain cool gas within the hot halo \citep{gronke18,fielding20}. In such environments, cooling gas may fragment into numerous small structures, motivating the ``misty'' CGM picture in which cool gas is distributed among many cloudlets rather than a few monolithic clouds \citep{mccourt18,gronke20}. 
In cosmological simulations, increasing CGM resolution generally produces smaller and more numerous cool structures and increases the predicted cool-gas covering fraction, bringing simulations into better agreement with observations \citep{voort19,nelson20,peeples19,ramesh24,augustin25}. 
These simulations also predict strong small-scale variability, such that neighbouring sightlines at similar impact parameters may intersect very different numbers of clouds. 
Within this context, the cloud size inferred from our beam-size analysis should be interpreted as an effective transverse coherence scale of the cloud complex rather than the diameter of an individual cloudlet. 
The beam-size dependence reported here directly links covering fraction, size, and transverse coherence, offering a promising way to connect multi-scale CGM structure predicted by theory with absorption-line observations.

At the current stage, the S/N and resolution of the stacked spectra, together with the limited leverage on the intrinsic background-light distribution, do not yet allow us to place tight quantitative constraints on the full set of toy-model parameters. In particular, parameters governing the cloud population and spatial distribution (e.g. $\alpha_m$, $\alpha_r$, $\eta$, $M_{\rm min}$, $M_{\rm max}$ and $r_{bg}$) remain degenerate when constrained only by a few data points of mean/median stacks. 
Nevertheless, when varying these parameters, the model produces distinct and testable signatures in both the absorption-strength profiles and the covering-fraction curves (Section~\ref{subsec_parameter_variation}). 
This highlights a promising path forward to the next generation of large spectroscopic datasets. Applying similar foreground--background stacking analyses to substantially larger samples, and crucially, to background sources spanning a wider range of intrinsic sizes and surface-brightness profiles, will enable tighter constraints on model parameters, and may allow environment, mass, and orientation effects to be separated rather than blended. In particular, forthcoming spectroscopic surveys on 10m-class telescopes \citep{maiolino20,greene22,mainieri24} will provide drastically increased numbers of high quality background spectra per foreground halo, sharpening the connection between beam averaging and absorber geometry and turning extended-source absorption from a qualitative diagnostic into a quantitative tool for mapping the small-scale structure of the cool CGM.
In parallel, comparisons with virtual observations generated from hydrodynamical simulations will be essential for constraining the multiphase CGM structure and the underlying microphysics, kinematics, and line-of-sight projection effects.

\section{Conclusions}
\label{sec_conclusion}

The characteristic scale of cool CGM structures remains difficult to constrain observationally. 
In this paper, we have introduced a new approach that compares stacked \mgii\ absorption measured against \emph{extended} background galaxies with that measured against effectively pencil-beam background quasars. 
This comparison provides a statistical and model-independent way to constrain the characteristic coherence scale of cool \mgii-bearing gas. 

Using the MEGAFLOW MUSE dataset, we constructed a large (5503) foreground--background pair catalogue and extracted continuum-normalised spectra for both background galaxies and background quasars. 
The foreground galaxies are at redshift $z\approx 1$, with a median stellar mass $\log_{10}(M_\star/M_\odot)=9.216$.
The background galaxies have a median radius $\approx$2.1\,kpc and a median continuum magnitude $\approx$25.7\,mag.
Stacking these spectra in bins of projected separation, $b/R_{\rm vir,fg}$, we computed the mean and median \mgii\ absorption as a function of the projected separation.

Our main observational result is that the \emph{mean--median contrast} differs strongly between background sources (Figure~\ref{fig_EW1}). 
For quasar sightlines, the median \mgii\ EW declines much more steeply with radius than the mean, and the difference becomes increasingly pronounced at large separations, reaching $S\simeq10$ in the outermost bin. By contrast, the galaxy-background stacks show much closer agreement between the mean and median, with $S\lesssim1.5$ in all bins. 
This behaviour is naturally explained if \mgii\ absorption arises in clumpy, partially covering structures (Figure~\ref{fig_schematic_intro}): pencil-beam quasars are highly sensitive to whether a sightline intersects a cloud, whereas extended background galaxies average over many neighbouring sub-sightlines and therefore smooth out the strong sightline-to-sightline variance seen in the quasar case.

In addition, the mean and median EW radial profiles differ when the background-galaxy sample is split by size into large (\(r_{\rm bg}\approx 6.6\)~kpc) and small (\(r_{\rm bg}\approx 1.5\)~kpc) subsamples. 
Although these measurements are noisier than those for the full sample, 
the small-background-galaxy stacks show a tentative mean--median separation in the intermediate radial bin ($S\simeq3$), similar to the trend seen for quasars, whereas the large-background-galaxy stacks remain broadly consistent between the two estimators.
This size dependence strengthens the beam-size interpretation and suggests that the effective transverse coherence scale of the foreground cool gas is on kpc scales, comparable to the background-galaxy sizes ($\approx2-7$~kpc) probed here.

To interpret these trends, we developed a simple ``cloud-in-halo'' toy model whose purpose is not to reproduce the full thermal/ionisation structure of the CGM, but to test whether partial covering and beam averaging can account for the observations. The model qualitatively reproduces the key behaviour seen in the stacks (Figure~\ref{fig_model_profiles2}), linking the observed mean--median contrast to the covering statistics and effective sizes of \mgii-bearing structures. In this sense, our results support a picture in which the cool CGM exhibits transverse coherence on scales of order kpc.

\begin{acknowledgements}
This work has been carried out thanks to the support of
the ANR 3DGasFlows (ANR-17-CE31-0017) and the ANR DARK (ANR-22-CE31-0006). The data used in this work are based on observations made with ESO telescopes at the La Silla Paranal Observatory and available in the ESO
archive (http://archive.eso.org). The MEGAFLOW catalogs and reduced cubes are available at https://megaflow.univ-lyon1.fr/ and https://amused.univ-lyon1.fr/megaflow/ where the catalog can be queried interactively.
RA acknowledges funding from the European Research Council (ERC) under the European Union's Horizon 2020 research and innovation programme (grant agreement 101020943, SPECMAP-CGM). 
\end{acknowledgements}

\bibliographystyle{aa} 
\bibliography{main} 

\begin{appendix}

\onecolumn
\section{Equivalent-width measurements}
\label{app_ew_table}

Table~\ref{tab:mean_median_ew} summarises the stacked
\mgii\ $\lambda2796$ EW measurements for the full sample and for the subsamples split by background-galaxy (BG) size and foreground-galaxy (FG) stellar mass. 

\begin{table*}[]
\centering
\tiny
\setlength{\tabcolsep}{4pt}
\caption{
Summary of stacked Mg\,\textsc{ii} $\lambda2796$ equivalent-width measurements.}
\label{tab:mean_median_ew}
\begin{tabular}{lllccccc}
\toprule[1.2pt]
Sample & BG type & Statistic & $r_{\rm bg}$ & FG mass
& $EW_{0-0.25R_{\rm vir}}$ & $EW_{0.25-1R_{\rm vir}}$ & $EW_{1-4R_{\rm vir}}$ \\
 & & & (kpc) & $\log_{10}(M_\star/M_\odot)$
& (\AA) & (\AA) & (\AA) \\
\midrule[1.2pt]

\multirow{4}{*}{Full sample}
& \multirow{2}{*}{Galaxy} & Mean   & \multirow{2}{*}{2.1} & \multirow{2}{*}{9.216}
& $1.026\pm0.116$ & $0.352\pm0.036$ & $0.190\pm0.026$ \\
&                         & Median &                      &
& $1.111\pm0.156$ & $0.263\pm0.045$ & $0.130\pm0.030$ \\

& \multirow{2}{*}{Quasar} & Mean   & \multirow{2}{*}{$\ll 10^{-5}$} & \multirow{2}{*}{9.216}
& $1.496\pm0.134$ & $0.619\pm0.054$ & $0.292\pm0.029$ \\
&                         & Median &                         &
& $1.169\pm0.169$ & $0.127\pm0.042$ & $0.014\pm0.004$ \\

\midrule[1.2pt]

\multirow{4}{*}{Different BG sizes}
& \multirow{2}{*}{Large galaxy} & Mean   & \multirow{2}{*}{6.6} & \multirow{2}{*}{9.212}
& $1.791\pm0.666$ & $0.604\pm0.183$ & -- \\
&                               & Median &                      &
& $1.847\pm0.818$ & $0.586\pm0.238$ & -- \\

& \multirow{2}{*}{Small galaxy} & Mean   & \multirow{2}{*}{1.5} & \multirow{2}{*}{9.230}
& $1.133\pm0.189$ & $0.358\pm0.069$ & -- \\
&                               & Median &                      &
& $1.057\pm0.274$ & $0.063\pm0.075$ & -- \\

\midrule[1.2pt]

\multirow{6}{*}{Different FG masses}
& \multirow{2}{*}{Galaxy} & Mean   & \multirow{2}{*}{2.1} & \multirow{2}{*}{$>9.216$}
& $1.272\pm0.127$ & $0.401\pm0.052$ & -- \\
&                         & Median &                      &
& $1.347\pm0.194$ & $0.303\pm0.065$ & -- \\

& \multirow{2}{*}{Quasar} & Mean   & \multirow{2}{*}{$\ll 10^{-5}$} & \multirow{2}{*}{$>9.216$}
& $1.547\pm0.168$ & $0.662\pm0.067$ & $0.335\pm0.044$ \\
&                         & Median &                         &
& $1.148\pm0.243$ & $0.175\pm0.065$ & $0.022\pm0.007$ \\

\cmidrule(lr){2-8}

& \multirow{2}{*}{Quasar} & Mean   & \multirow{2}{*}{$\ll 10^{-5}$} & \multirow{2}{*}{$<9.216$}
& $1.374\pm0.236$ & $0.550\pm0.091$ & $0.258\pm0.035$ \\
&                         & Median &                         &
& $1.150\pm0.328$ & $0.064\pm0.050$ & $0.009\pm0.004$ \\

\bottomrule[1.2pt]
\end{tabular}
\end{table*}

\onecolumn
\section{The \mgii\ absorption variation with background galaxy size}
\label{app_size_dependence}

We explore whether the stacked \mgii\ absorption profiles depend on the effective size of the BG galaxies. 
To do so, we split the BG-galaxy sample into two subsamples using the 75th-percentile value of the effective BG radius. 
We refer to these two subsamples as ``large'' (median $r_{bg}\approx$6.6~kpc) and ``small'' BG (median $r_{bg}\approx$1.5~kpc) galaxies. 
The small-BG stacks contain more spectra than the large-BG stacks, because larger BG galaxies are more likely to be brighter and have higher continuum S/N. 
The median FG stellar masses of the large- and small-BG subsamples are $\log_{10}(M_\star/M_\odot)=9.21$ and 9.23, respectively.
The spectra are then stacked in the same three bins of normalised separation as in the main text, $b/R_{\rm vir,fg} \in [0,0.25)$, $[0.25,1)$, and $[1,4)$, using the same mean/median estimators and bootstrap uncertainty calculation as described in Section~\ref{subsec_stacking}.
The corresponding EW measurements are listed in Table~\ref{tab:mean_median_ew}.

\begin{figure}
    \centering
    \includegraphics[width=.7\columnwidth]{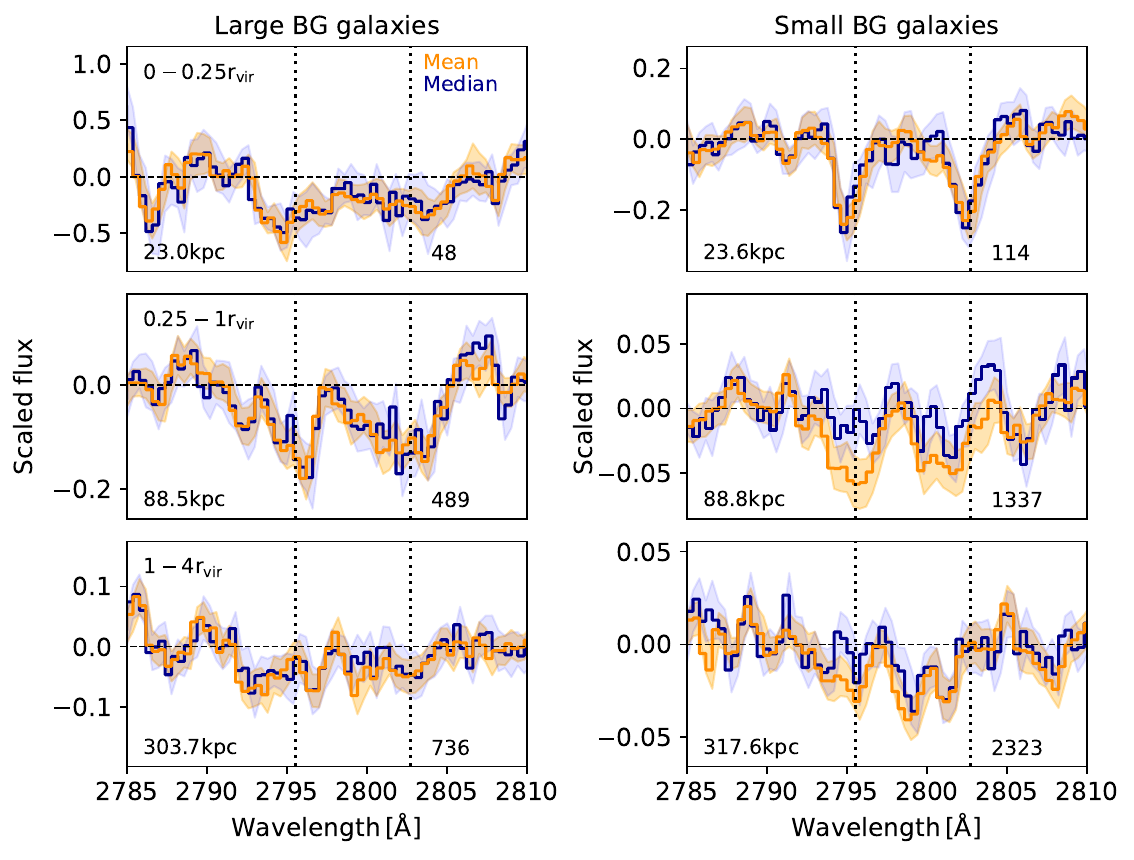}
    \caption{
    Same as Figure~\ref{fig_stack_spec1}, but for the large and small background-galaxy subsamples.
    }
    \label{fig_stack_spec_size1}
\end{figure}

Figure~\ref{fig_stack_spec_size1} shows the stacked \mgii\ absorption profiles for the two BG-size subsamples. 
The left column presents sightlines traced by large BG galaxies, while the right column shows those traced by small BG galaxies. 
In both subsamples, the \mgii\ doublet is most clearly detected in the inner and intermediate radial bins, and becomes largely undetectable at large radii. 
Figure~\ref{fig_EW_bins_size1} shows the \mgii\ EW as a function of normalised separation for the large- and small-BG subsamples, together with the quasar measurements from the main text for reference. 
From the inner to the intermediate radial bins, the small-BG subsample shows a systematic separation between the mean and median measurements, whereas no clear mean--median difference is detected in the large-BG stacks. We caution, however, that the size-split measurements are noisier than those of the full sample. In particular, for the small-BG subsample in the intermediate radial bin, the mean stack yields an \mgii\ detection at the $\approx 5\sigma$ level, while the median stack is detected only at the $\approx 1\sigma$ level.
The outermost BG-galaxy bin is not shown in this figure because the stacked absorption becomes too weak for a robust size-dependent comparison.

These trends are qualitatively consistent with the geometric interpretation developed in the main text (Figure~\ref{fig_schematic_intro}). 
If the cool \mgii$-$bearing gas is composed of clouds and has typical covering on kpc scales, then the observed absorption against a background galaxy depends not only on whether the foreground clouds intersect the line of sight, but also on how the cloud distribution is averaged over the projected extent of the background light. 
A larger BG galaxy samples a wider bundle of sub-sightlines and therefore has a higher probability of intersecting at least part of the cloud population. 
This can increase the effective covering fraction and maintain a detectable absorption signal even when individual sub-sightlines would otherwise miss the clouds. 
At the same time, because the absorption is averaged over a larger beam, the mean and median remain relatively close, rather than developing the stronger separation seen for the small-BG sample.

Overall, splitting the sample by BG galaxy size strengthens our interpretation: not only do quasars and galaxies show different mean--median behaviour, but even within the BG-galaxy sample there are indications that background beams of different sizes produce different stacked absorption strengths. 
This is exactly the behaviour expected if the cool CGM is structured on scales comparable to the size range of galaxies, so that partial covering and beam averaging both contribute to shaping the observed stacked \mgii\ profiles.

\section{The \mgii\ absorption variation with foreground galaxy mass}
\label{app_mass_dependence}

\begin{figure*}
    \centering
    \includegraphics[width=0.49\textwidth]{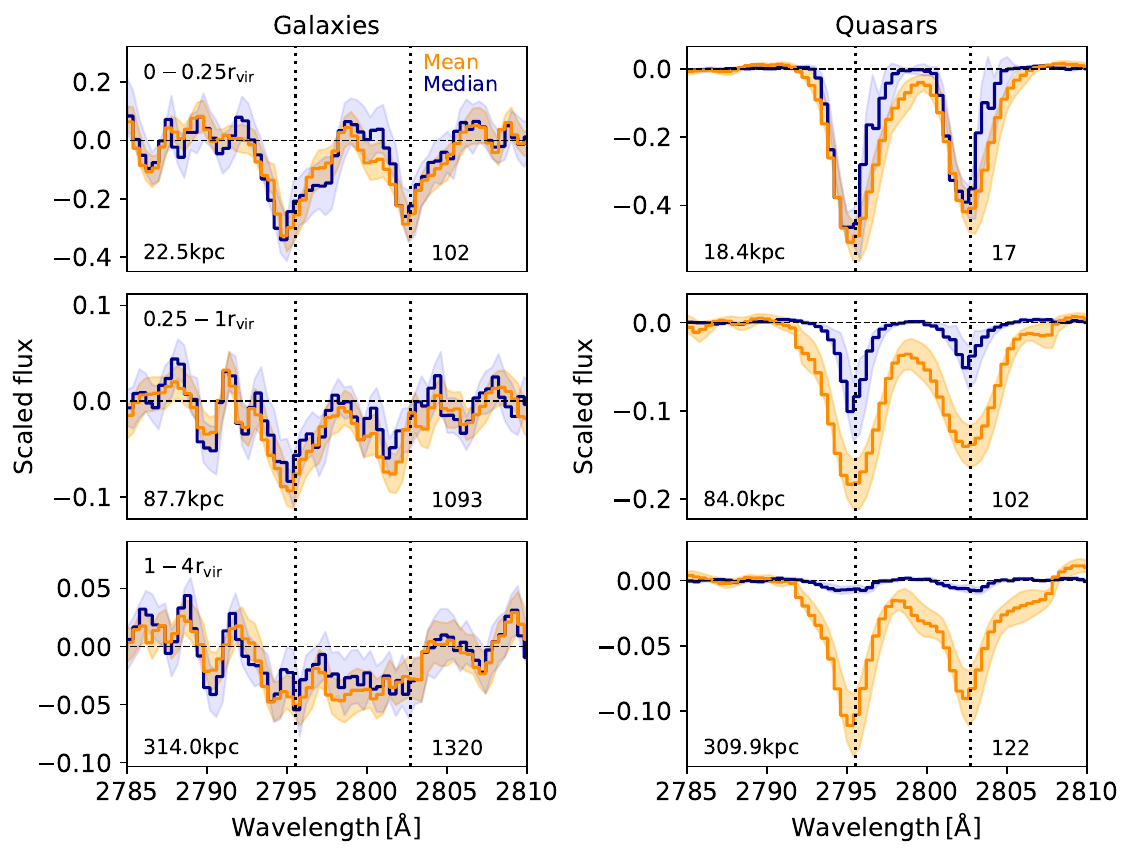}
    \hfill
    \includegraphics[width=0.49\textwidth]{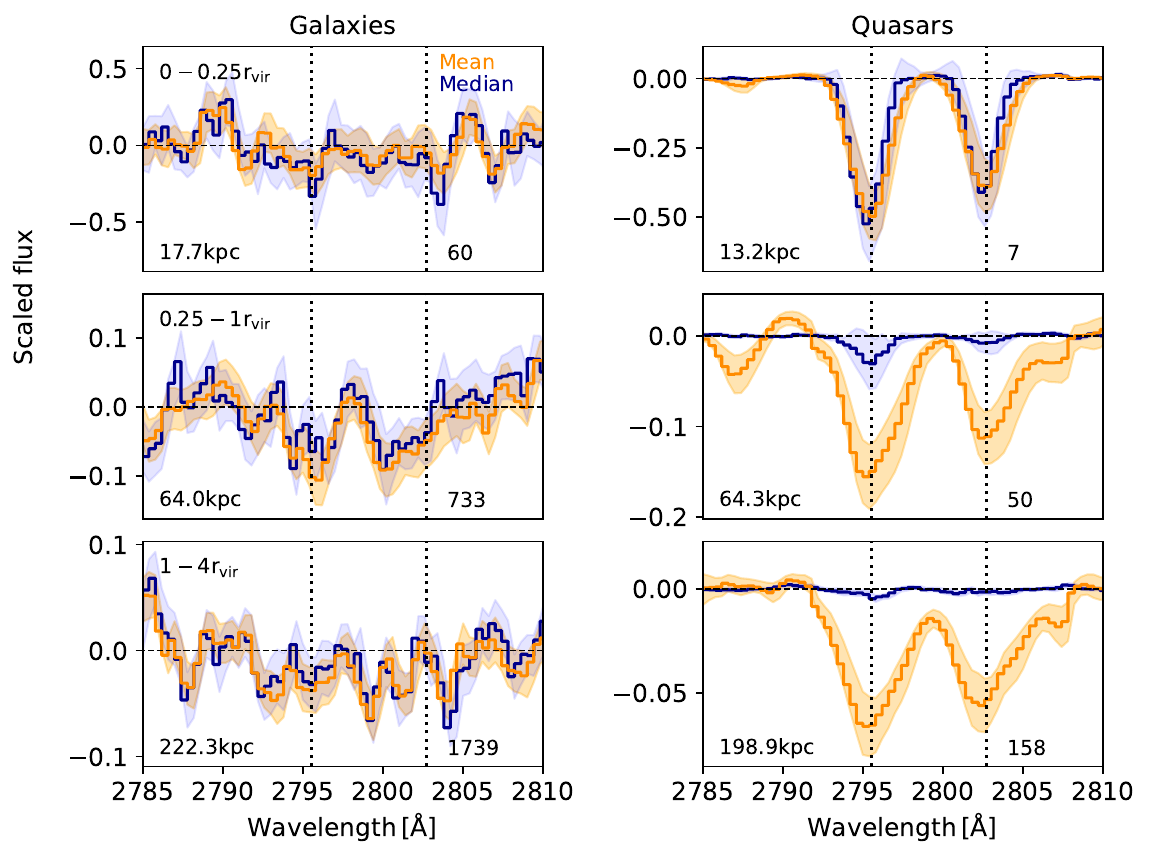}
    \caption{
    Same as Figure~\ref{fig_stack_spec1}, but for the foreground-galaxy mass subsamples.
    Left: the high-mass foreground-galaxy subsample (($\log_{10}(M_\star/M_\odot)> 9.216$)).
    Right: the low-mass foreground-galaxy subsample ($\log_{10}(M_\star/M_\odot)< 9.216$).
    }
    \label{fig_stack_spec_mass1}
\end{figure*}

\begin{figure}
    \centering
    \includegraphics[width=0.45\columnwidth]{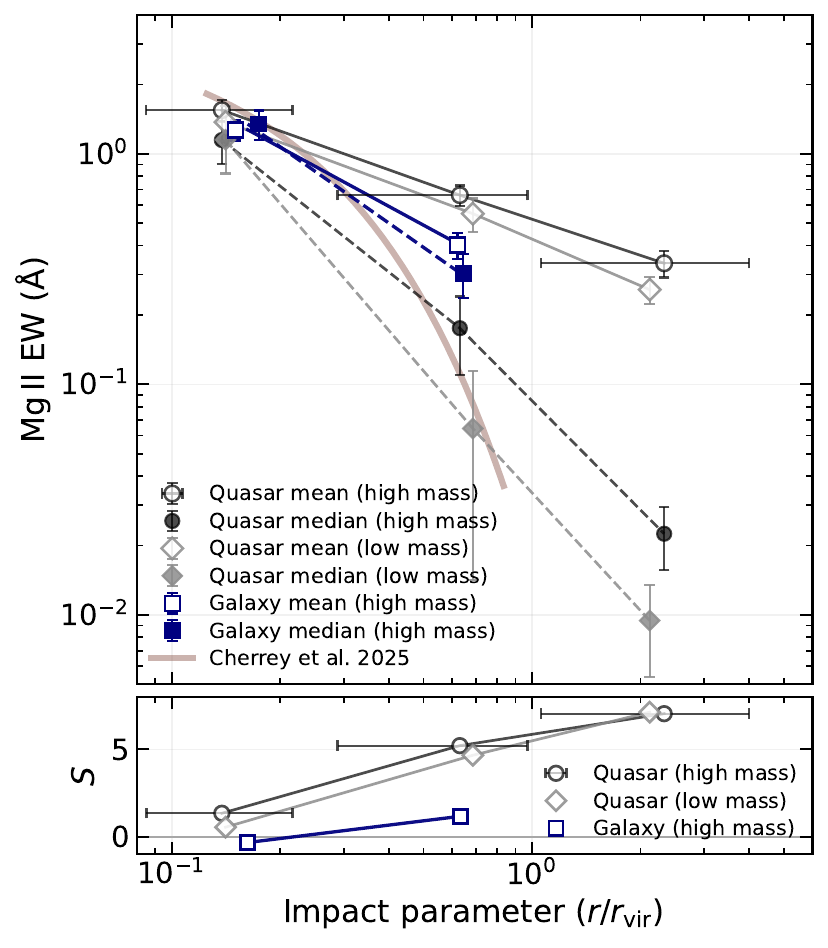}
    \caption{
    Same as Figure~\ref{fig_EW1}, but with the sample split into low- and high-mass foreground galaxies. The top panel shows the radial \mgiiaw\ EW profiles from the mean and median stacks. The bottom panel shows the corresponding mean--median contrast significance, $S$. The quasar stacks show a clear mean--median contrast in both foreground-mass bins, whereas the high-mass foreground-galaxy stack measured against background galaxies shows a much smaller contrast. 
    For the high-mass background-galaxy subsample, the outermost radial bin is omitted because the S/N is insufficient for a reliable EW measurement. For the low-mass background-galaxy subsample, the \mgii\ absorption is too weak to yield reliable EW constraints, and these measurements are therefore not shown.
    }
    \label{fig_EW_bins1}
\end{figure}

We explore whether the stacked \mgii\ absorption profiles depend on the stellar mass of the FG galaxies. We split the FG--BG pair catalogue into two equal-sized subsamples using the median foreground stellar mass derived from the SED-fitting products ($\log_{10}(M_\star/M_\odot)=9.216$ in our working catalogue). 
The two mass subsamples are then stacked in the same three radial bins of normalised separation as in the main text. 

The left panel of Figure~\ref{fig_stack_spec_mass1} shows the stacked profiles for the high-mass subsample. The \mgii\ doublet is clearly detected in the BG-galaxy stacks at small and intermediate separations. 
The right panel of Figure~\ref{fig_stack_spec_mass1} shows the corresponding low-mass stacks. In the BG-galaxy stacks, \mgii\ absorption is only marginally detected in the intermediate radial bin ($0.25<R/R_{\rm vir,fg}<1$), and is not convincingly detected in either the innermost or the outermost bins. 
This behaviour likely reflects a combination of intrinsically less cold gas in the CGM of low-mass galaxies and reduced S/N.
In addition, the lack of a clear detection at the smallest separations may have a distinct physical origin: at small impact parameters the CGM can exhibit significant \mgii\ emission, which can fill in the absorption trough \citep[e.g.][]{guo23}. This effect is expected to be particularly important for galaxy-BG spectra, which are surface-brightness--weighted averages over an extended, typically faint continuum source; any localised emission and/or uncovered bright regions can therefore contribute residual flux that fills in the line and further suppresses the observed absorption depth, making detection more difficult.

We measure the EW from the mass-split stacked spectra and show the results in Figure~\ref{fig_EW_bins1}. The quasar-BG measurements are shown for both the high- and low-mass subsamples, while only the high-mass galaxy-BG measurements are plotted; for the low-mass galaxy-BG stacks the \mgii\ S/N is too low for reliable EW constraints in all three bins. 
The corresponding EW measurements are listed in Table~\ref{tab:mean_median_ew}.

Despite these mass-dependent differences in detectability, our main qualitative conclusions remain unchanged. In both mass bins, the quasar stacks exhibit a clear separation between the mean and median measurements, consistent with substantial sightline-to-sightline variance in pencil-beam absorption. 
Conversely, the galaxy-BG stacks (where measurable) show much closer agreement between mean and median, consistent with beam averaging across an extended BG light distribution.
Thus, the distinctive mean--median behaviour that motivates our partial-covering/beam-averaging interpretation is robust to splitting the sample by foreground stellar mass.

\section{The effect of noise on the mean-median contrast}
\label{app_noise}

Median stacks are more sensitive to the S/N of the input spectra than mean stacks (e.g. \citealt{mishra24}). For a skewed distribution in which most sightlines lie close to the continuum while a minority show strong absorption, the mean is pulled toward the absorbing tail, whereas the median remains close to the continuum unless absorption is present in more than half of the spectra. If the spectra are dominated instead by large, approximately symmetric noise, the pixel-flux distribution is broadened and some pixels are scattered below the continuum. This can increase the apparent median EW and reduce the mean--median contrast.

We therefore test whether lowering the S/N of the quasar spectra can remove the quasar mean--median contrast. 
To do this, we derive the distribution of continuum uncertainties in the background-galaxy spectra. For each quasar spectrum, we randomly assign one uncertainty drawn from this distribution and add Gaussian pixel noise with the corresponding amplitude to the quasar spectrum. We then stack the modified quasar spectra in the same way as for the fiducial stacks. This experiment therefore preserves the quasar absorption signal while imposing a realistic distribution of background-galaxy noise levels.
The median continuum-flux uncertainty of the background-galaxy spectra is $\sigma_{\lambda}\simeq9.3\times10^{-21}\,{\rm erg\,s^{-1}\,cm^{-2}\,\AA^{-1}}$.
We then repeat the full stacking and bootstrap analysis using the same radial bins and EW measurements as in the main text.

\begin{figure*}
    \centering
    \includegraphics[width=0.59\textwidth]{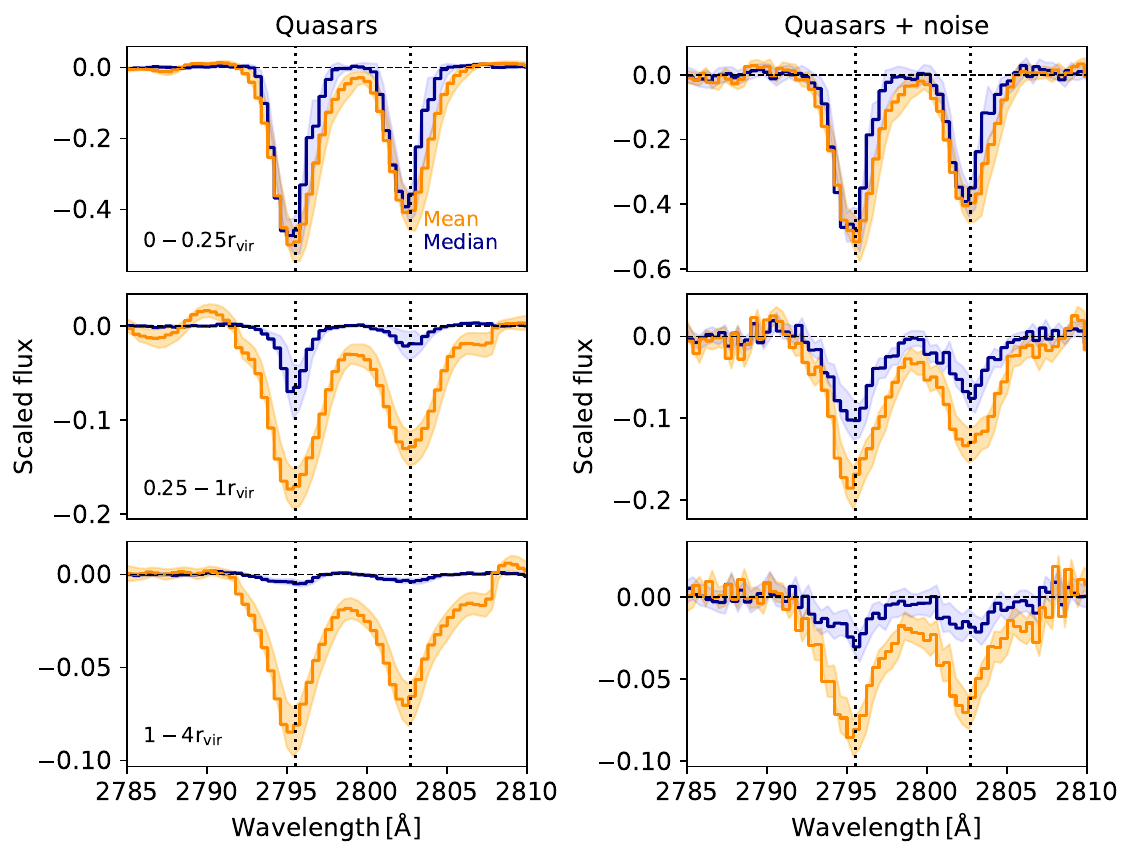}
    \hfill
    \includegraphics[width=0.39\textwidth]{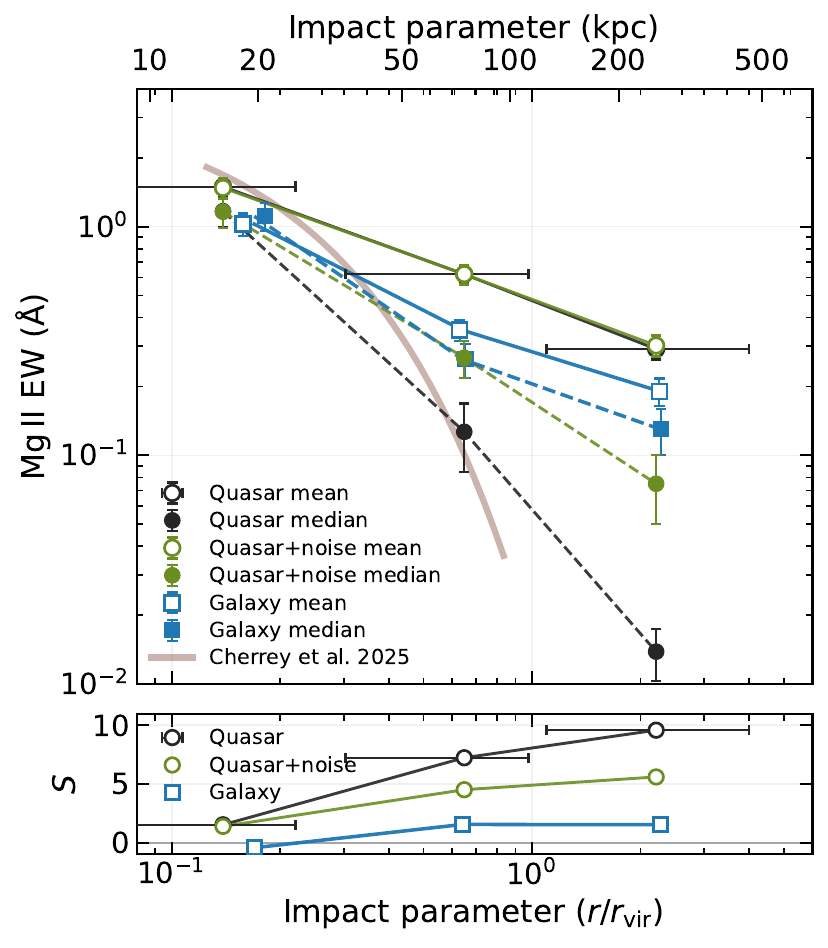}
    \caption{
    Left: stacked quasar spectra before and after adding noise. The first column shows the fiducial quasar stacks, while the second column shows the stacks obtained after adding Gaussian noise to each individual quasar spectrum, with the noise amplitude drawn from the observed distribution of background-galaxy continuum uncertainties.
    Right: same as Figure~\ref{fig_EW1}, but including the noise-added quasar measurements. At a noise level comparable to that of the BG-galaxy spectra, the quasar median EW declines more slowly with radius, but the mean--median contrast remains evident.
    }
    \label{fig_stack_qso_spec_noise}
\end{figure*}

The left subplot of Figure~\ref{fig_stack_qso_spec_noise} shows the resulting stacked spectra.
Adding noise has only a marginal effect on the mean stack. The median stack is more affected, especially at large radius where the intrinsic absorption is weak. The added noise broadens the flux distribution and increases the apparent median absorption. This effect is also seen in the EW profiles shown in the right subplot of Figure~\ref{fig_stack_qso_spec_noise}. The noise-added median EWs are larger than in the fiducial quasar stacks, but the quasar mean--median contrast remains clear and does not approach the much closer mean--median agreement seen in the BG-galaxy stacks.
We therefore conclude that noise can partly reduce the quasar mean--median contrast, but cannot erase it at the noise level relevant for this analysis.

\FloatBarrier 
\clearpage

\end{appendix}
\end{document}